\documentclass[prl,amsmath,amssymb,twocolumn, notitlepage]{revtex4-1}
\usepackage{graphicx}
\usepackage{dcolumn}
\usepackage{bm}
\usepackage{textcomp}
\usepackage{float}
\usepackage[colorlinks]{hyperref}
\usepackage{epstopdf}

\usepackage{amsmath, amssymb, amsfonts, bm}
\usepackage{latexsym}
\usepackage[protrusion=true,expansion=true]{microtype}
\usepackage{caption}
\usepackage{subcaption}
\newcommand{\fixme}[1]%
   {\begingroup{\color{blue}\it (FIXME: #1)}\endgroup}
\newcommand{\highlight}[1]%
   {\begingroup{\color{blue} #1}\endgroup}
\begin{document}

\title{Observation of the Bloch--Siegert shift in a driven quantum-to-classical transition}

\author{I.~Pietik\"ainen$^1$}
\author{S.~Danilin$^2$}
\author{K.~S. Kumar$^2$}
\author{A. Veps\"al\"ainen$^2$}
\author{D.~S. Golubev$^2$}
\author{J.~Tuorila$^{1,3}$}
\author{G.~S.~Paraoanu$^2$}

\affiliation{$^1$Nano and Molecular Materials Research Unit, University of Oulu, P.O. Box 3000, FI-90014, Finland}
\affiliation{$^2$Centre for Quantum Engineering and LTQ, Department of Applied Physics, Aalto University, P.O. Box 15100, FI-00076 Aalto, Finland}
\affiliation{$^3$COMP Centre of Excellence, Department of Applied Physics, Aalto University, P.O. Box 15100, FI-00076 Aalto, Finland}

\date{\today}

\begin{abstract}
We show that the counter-rotating terms of the dispersive qubit-cavity Rabi model can produce relatively large and nonmonotonic Bloch--Siegert shifts in the cavity frequency as the system is driven through a quantum-to-classical transition. Using a weak microwave probe tone, we demonstrate experimentally this effect by monitoring the resonance frequency of a microwave cavity coupled to a transmon and driven by a microwave field with varying power. In the weakly driven regime (quantum phase), the Bloch--Siegert shift appears as a small constant frequency shift, while for strong drive (classical phase) it presents an oscillatory behaviour as a function of the number of photons in the cavity.
The experimental results are in agreement with numerical simulations based on the quasienergy spectrum.
\end{abstract}

\maketitle

The Rabi Hamiltonian - describing a two-level system coupled to a cavity (resonator) mode - is a paradigmatic model in quantum physics. In the rotating-wave approximation (RWA) it leads to the well-known Jaynes--Cummings (JC) model. In the dispersive limit this model predicts the appearance of ac-Stark shifts in the energy levels of both the qubit and the cavity. The inclusion of counter-rotating terms produces an additional displacement of the energy levels. This Bloch--Siegert (BS) shift~\cite{BlochSiegert40} is usually very small on standard experimental platforms since it depends on the ratio between the coupling and the sum of the Larmor and cavity frequencies. Recently however, significant experimental effort has been put into increasing the coupling to values comparable with the Larmor frequency~\cite{Ciuti2006,Devoret2007,Bourassa2009}, most notable in semiconductor dots~\cite{Gunter2009,Todorov2010} and superconducting circuits~\cite{FornDiaz10,Niemczyk2010,FornDiaz16,Yoshihara16,Baust2016,Bishop09}. Other approaches for observing the Bloch--Siegert shift include the detailed analysis of two-level Landau--Zener spectra of Rydberg atoms~\cite{Fregenal04,Forre04} and Cooper-pair boxes~\cite{Goorden03,Tuorila10,Tuorila13}, as well as the simulation of the Rabi model in rotating frames~\cite{Li13,Braumuller16}.

\begin{figure}[th!]
\centering
\includegraphics[width=1\linewidth]{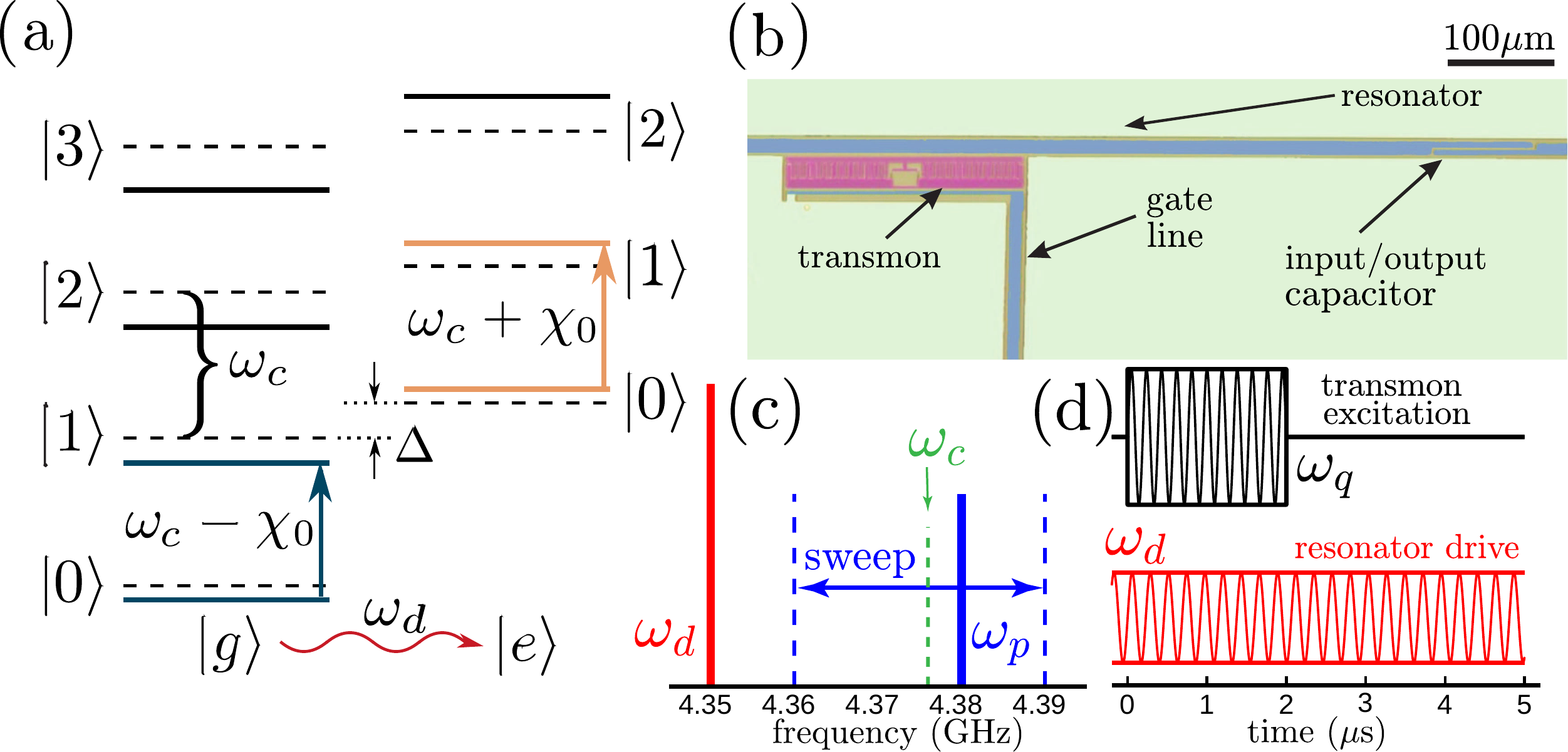}
\caption{Schematic of the experimental setup. (a) Energy levels of a dispersive JC system. The vacuum ac-Stark shift is given by $\chi_0=g^2/(\omega_0-\omega_{\rm c})$. (b) Optical micrograph of the main elements of the sample (false colors).
(c) We drive the cavity with a detuned drive frequency $\omega_{\rm d}<\omega_{\rm c}$. The spectrum  of the system is monitored by a weak probe with frequency $\omega_{\rm p}$, which is swept within the window $[4.36, 4.39]$ GHz. (d) In the $T_1$ measurements, the cavity is driven continuously at $\omega_{\rm d}$. In the steady state, a long $2 \mu$s microwave pulse at the ac-Stark shifted qubit frequency $\omega_q \gg \omega_{\rm c}$ is applied to the gate line of the qubit. When the pulse ends, the decay traces of the system relaxing back to the steady state are monitored in the time domain using a homodyne detection scheme.
} \label{fig:schema}
\end{figure}

In this work we take a different route. We recognize that the counter-rotating terms do not conserve the excitation number. 
Therefore, the natural framework for their experimental demonstration is that of driven-dissipative systems~\cite{Hartmann16,Noh16}. Guided by this intuition, we realize a setup consisting of a transmon~\cite{Koch07} dispersively coupled to a cavity, where the cavity is driven at a fixed off-resonance microwave tone, while at the same time the spectrum is scanned by a comparatively weaker probe field, see Fig.~\ref{fig:schema}. At low driving powers, we observe the expected vacuum ac-Stark shift~\cite{Reed10,Boissonneault10,Bishop10,Suri15, Paik2011}. This is followed by a transition regime dominated by nonlinear effects as the power is increased. For the Jaynes--Cummings model, such transition has been predicted and studied in the resonant qubit-cavity case~\cite{Alsing92,Carmichael15,Fink10,Fink17}. In the dispersive limit, the transition region is no longer abrupt, but it is expected to soften into a Kerr-type nonlinearity~\cite{Carmichael15}, in agreement with our observations.

As the system approaches and enters the classical phase, the frequency renormalization due to the BS effect 
results in a strongly nonmonotonic dependence on the number of photons in the cavity. The effect escapes the scope of the simple RWA, but can be modeled numerically using the Floquet formalism~\cite{Shirley65} for the driven cavity coupled with a multilevel transmon. The measured deviation from the RWA result is in agreement with the simulations and constitutes the first experimental observation of the BS effect in a driven quantum-to-classical transition. 

{\em Theoretical predictions for a two-level system -- }
We develop a simple physical picture of these phenomena by considering the two lowest transmon states coupled with the cavity which is driven by a tone of amplitude $A$ and frequency $\omega_d$. The resulting driven Rabi Hamiltonian (scaled with $\hbar$) is written as
\begin{equation}\label{eq:driven RabiHam}
\hat{H}= \omega_{\rm c} \hat{a}^{\dag}\hat{a} + \frac{\omega_0}{2}\hat{\sigma}_{\rm z} + g(\hat{a}^{\dag}+\hat{a})\hat{\sigma}_{\rm x} + A\cos(\omega_{\rm d} t)(\hat{a}^{\dag}+\hat{a}),
\end{equation}
where $\omega_{\rm c}$ and $\omega_{0}$ are the frequencies of the cavity and the qubit, respectively, and $g$ is the strength of the coupling, see Fig.~\ref{fig:schema}. We displace the Hamiltonian in Eq.~(\ref{eq:driven RabiHam}) into the vacuum and apply the RWA. After a Schrieffer--Wolff transformation, we obtain the driven JC Hamiltonian~\cite{supplement}
\begin{eqnarray}\label{eq:diagHam}
\hat{H} &=& (\omega_{\rm c} + \chi_0\hat{\sigma}_{\rm z})\hat{a}^{\dag}\hat{a} + \frac{1}{2}(\omega_0+\chi_0)\hat{\sigma}_{\rm z}\nonumber\\
&& + \frac{G}{2}\left(e^{i\omega_{\rm d} t}\hat{\sigma}_- + e^{-i\omega_{\rm d} t}\hat{\sigma}_+\right), \label{eq:RWAtwolevel}
\end{eqnarray}
where $\chi_{0} = g^2/(\omega_{0}-\omega_{\rm c})$ is the vacuum ac-Stark shift. Thus, the cavity frequency $\omega_{\rm c} + \chi_0\hat{\sigma}_{\rm z}$ depends on the qubit state, which is modulated by the 
the drive with the effective strength $G=gA/\sqrt{(\omega_{\rm c}-\omega_{\rm d})^2+\kappa^2/4}$ where $\kappa$ is the cavity dissipation. Consequently at high powers, the effective cavity frequency becomes a weighted average over the qubit state~\cite{Abragam,Li13}, i.e. simply $\omega_{\rm c}$. 

However, $G$ diverges with increasing $A$, thus we expect that the RWA is not valid at large drive powers. To capture the behaviour in this regime, we again transform the cavity in Eq.~(\ref{eq:driven RabiHam}) into the vacuum but then apply the counter-rotating hybridized rotating-wave (CHRW) approximation~\cite{Lu12,Yan15}. After a Schrieffer--Wolff transformation, we obtain Eq.~(\ref{eq:RWAtwolevel}), but with the renormalized qubit frequency $\tilde{\omega}_0\equiv\omega_0J_0(2G\xi/\omega_{\rm d})$ and drive amplitude $\tilde{G}/2 \equiv G(1-\xi)$. The parameter $\xi$ is determined by solving $G(1-\xi)=\omega_0 J_1(2G\xi/\omega_{\rm d})$~\cite{supplement}.

\begin{figure}[t]
\centering
\includegraphics[width=1.0\linewidth]{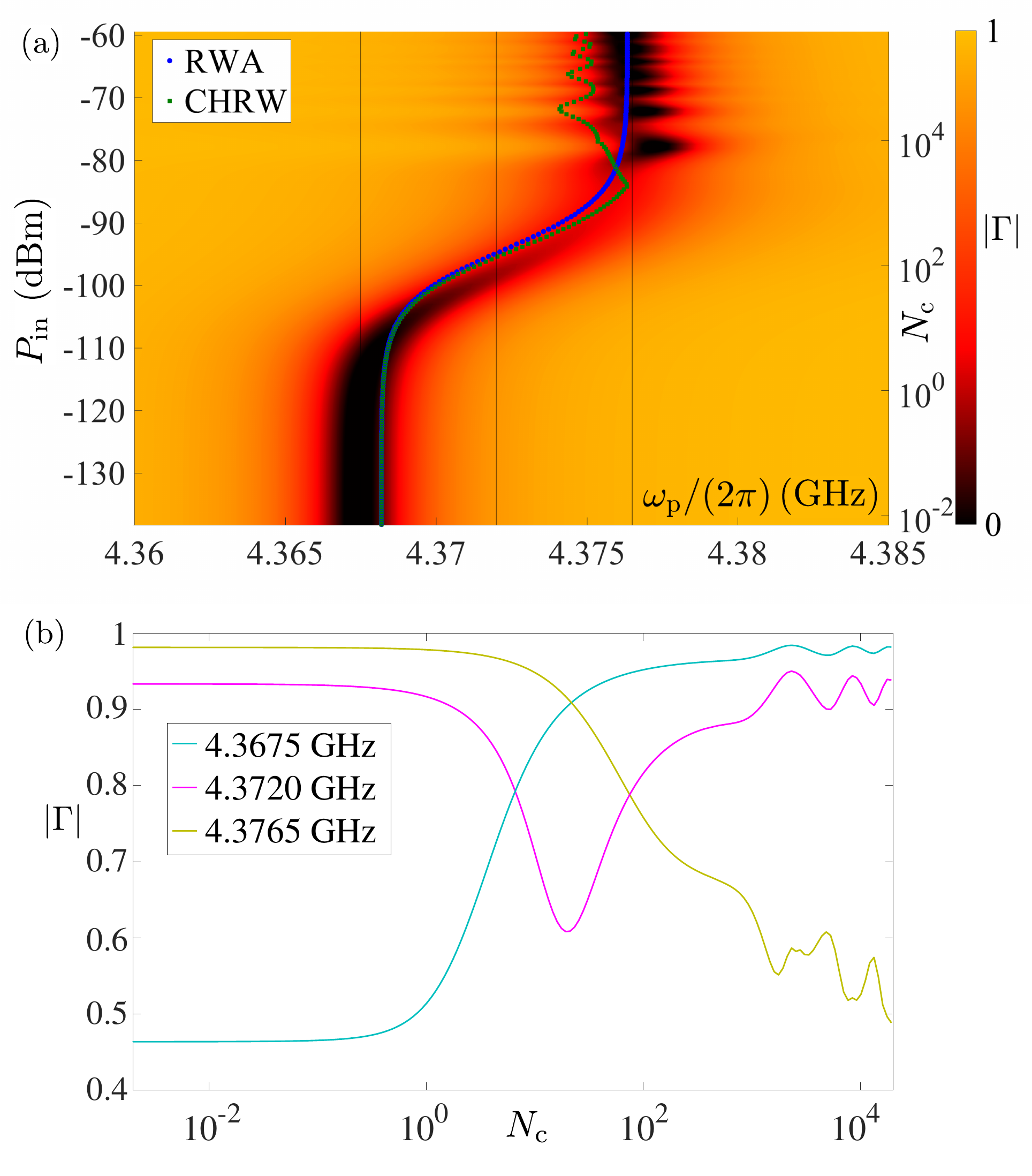}
\caption{Numerical simulation of the spectrum for a two-level system, corresponding to experimental values $\omega_{\rm c}/(2\pi)=4.376$ GHz, $\omega_0/(2\pi)=5.16$ GHz, $g/(2\pi)=80$ MHz, $\omega_{\rm d}/(2\pi) = 4.350$ GHz, and $\kappa/(2\pi) = 4.0$ MHz. (a) Probe reflection coefficient $|\Gamma|$ as a function of probe frequency $\omega_{\rm p}$ and cavity occupation $N_{\rm c}$. The locations of resonance in the RWA (blue) and in the CHRW (green) approximation are shown as well. (b) Reflection coefficients for the three different probe frequencies indicated in (a) with dashed vertical lines.}\label{fig:BSshift}
\end{figure}

We show the effective cavity resonance frequencies in the RWA and CHRW approximation in Fig.~\ref{fig:BSshift}(a). We compare the results with the numerical simulations~\cite{supplement} of the probe reflection coefficient $\Gamma(\omega_{\rm p})=(Z(\omega_{\rm p})-Z_0)/(Z(\omega_{\rm p})+Z_0)$, where $Z(\omega_{\rm p})$ and $Z_0$ are the impedances of the driven cavity coupled to a two-level transmon and the transmission line, respectively. The impedance $Z(\omega_{\rm p})$ was calculated from the probe induced transition rates between the quasienergy states of the driven cavity-transmon system by employing the Kramers-Kronig relation~\cite{Silveri13,Tuorila13,LL,supplement,Blumel89,Blumel91,Grifoni98,Gasparinetti13,Hausinger11, HausingerPhD}. We see that the RWA gives the overall qualitative behaviour relatively well but lacks the nonmonotonic behaviour of the simulated resonance frequency toward the high-power end of the spectrum. The CHRW approximation is nonmonotonic and is quite accurate when compared to the full numerical solution with $G/\omega_{\rm d} < 1$, and even for higher values of $G$ when $\omega_0/\omega_{\rm d} <1$~\cite{Lu12}. However, it still fails to match exactly the average numerical resonance at high drive powers, since it neglects the second and higher harmonics of the drive. 
At low powers, the deviation between the analytic and numerical resonance locations is caused by the vacuum BS shift $\chi_{\rm BS}=g^2/(\omega_0+\omega_{\rm c})=2\pi \times 0.7$ MHz. Strong driving clearly amplifies the BS effect, which appears as a deviation of nearly 3 MHz between the RWA and CHRW reflection minima. 

From Fig.~\ref{fig:BSshift}(b) we can see the existence of three regimes: at small average photon numbers, the response of the system is quantum-mechanical and corresponds to a cavity with constant frequency shift. When the number of photons becomes of the order of unity, the response is sensitive to the addition or removal of photons, indicating a dispersive photon blockade~\cite{Houck2011}. At large number of photons, the photon blockade breaks~\cite{Carmichael15}, and the Bloch--Siegert shift produces an oscillatory response as a function of added photons.

{\em Experimental results --} The physical device [Fig.~\ref{fig:schema}(b)] consists of a 
$\lambda/4$-waveguide-resonator cavity capacitively coupled to a transmon qubit with $E_{\rm J\Sigma}/E_{\rm C}=58$, flux-biased at $\omega_0/(2\pi)=5.16$ GHz. For the cavity, $\omega_{\rm c}/(2\pi)=4.376$ GHz and $\kappa/(2\pi) = 4.0$ MHz. The signals used for spectroscopy and for relaxation measurements are shown in Fig.~\ref{fig:schema}(c),(d).

\begin{figure}[h!]
\centering
\includegraphics[width=1.0\linewidth]{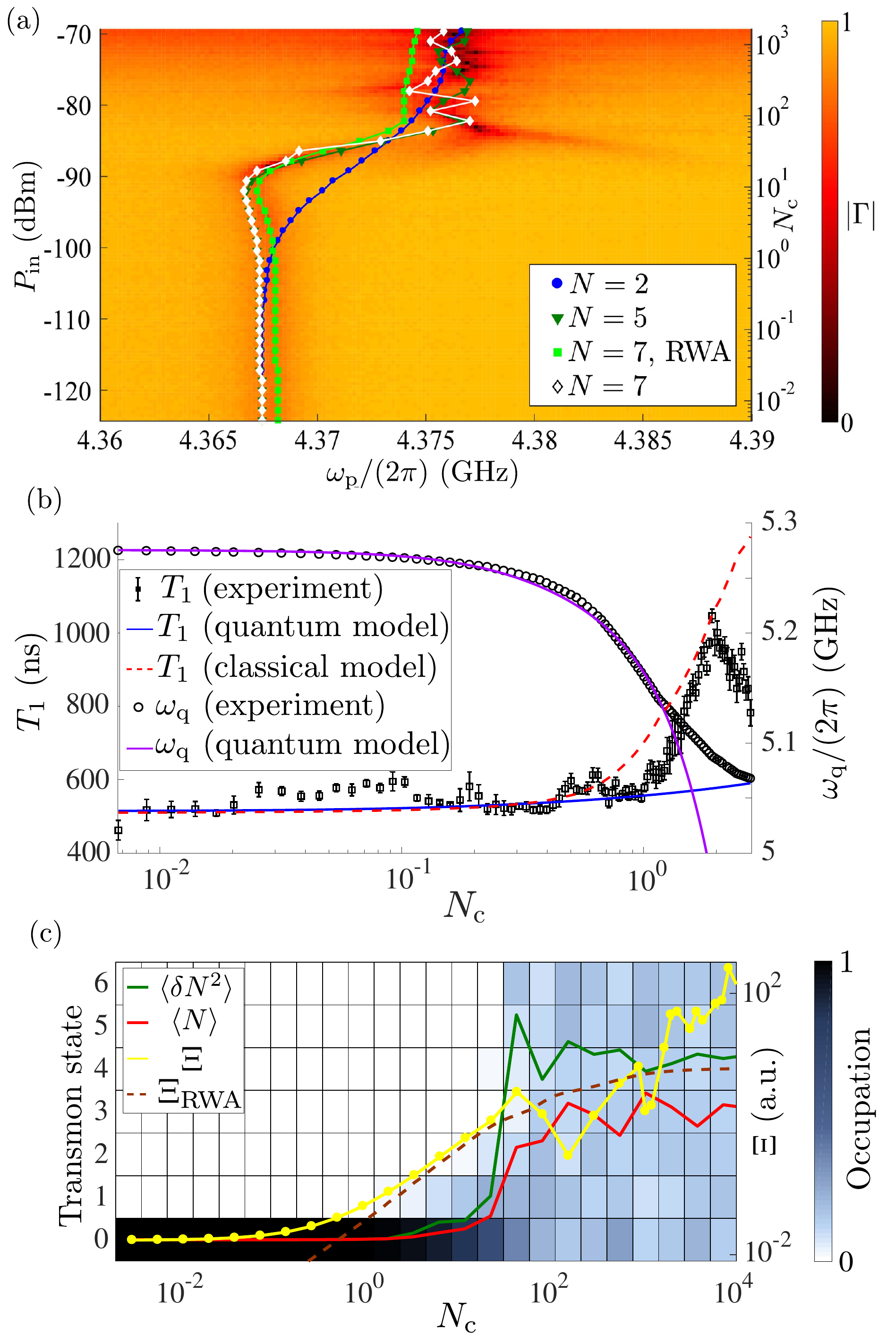}
\caption{Frequency renormalization and Bloch--Siegert effect. (a) Measured reflection coefficient $|\Gamma|$ as a function of the probe frequency $\omega_{\rm p}$ and the input power $P_{\rm in}$ at the sample (or equivalent average number of cavity quanta $N_{\rm c}$). Numerically-calculated resonance frequencies are shown with coloured dots for $N=2,5,7$ transmon states. (b) Measured relaxation time $T_1$ (left axis) as a function of $N_{\rm c}$. On the right axis, we show the ac-Stark shifted qubit transition frequency $\omega_{\rm q}$ in the same interval. For $N_{c}\leq 1$, a standard perturbative quantum treatment \cite{Boissonneault09,Boissonneault12a,Boissonneault12b} can be used to explain the data (continuous lines), while above these values the increase of $T_1$ can be modelled using a classical approach~\cite{supplement,Irwin,Irwin2}. 
(c) Quantum-to-classical transition in the driven transmon-cavity system. We show the numerically calculated average transmon state occupation $\langle N\rangle$, occupation number fluctuations $\langle \delta N^2\rangle$, and the order parameter $\Xi$ as a function of the cavity occupation.
}\label{fig:expBSshift}
\end{figure}

In Fig.~\ref{fig:expBSshift}(a), we present the measured reflection coefficient $\Gamma$ as a function of the probe frequency $\omega_{\rm p}$ and the number of cavity quanta $N_{\rm c}$ at the off-resonant drive frequency $\omega_{\rm d}/(2\pi) = 4.35$ GHz. The spectrum clearly shows the quantum-to-classical transition and the BS effect, as expected from the previous theoretical considerations and they are well-matched by numerical simulations for the Hamiltonian in Eq.~(\ref{eq:driven RabiHam}), generalized to the case of a multilevel transmon~\cite{Koch07,CottetThesis,supplement} and including the counter-rotating terms. The numerics have converged up to $N_{\rm c} \approx 100$ for $N=7$ transmon states. We have also performed a systematic experimental study of the spectrum at various transmon and drive frequencies~\cite{supplement}, obtaining similar features to those presented in Fig.~\ref{fig:expBSshift}(a).

In the quantum regime, the transmon behaves as a two-level system (qubit) with only $\hat{\sigma}_{\rm z}$-coupling to the cavity via a $\hat{a}^{\dag} \hat{a} \hat{\sigma}_z$-term, cf. Eq.~(\ref{eq:diagHam}). Consequently, the two systems can be addressed separately and, because $[\hat{H},\hat{\sigma}_{\rm z}] = 0$, the cavity can be used for quantum nondemolition measurements of the qubit population. The higher excited states of the transmon start to contribute to the observed resonance location when $N_{\rm c}\sim 1$, as seen from the deviation between the numerical $N=2$ and $N=7$ results in Fig.~\ref{fig:BSshift}(a). 
This coincides with the breakdown of the dispersive approximation used in Eq.~(\ref{eq:RWAtwolevel}), which is expected to occur when $N_{\rm c}\ll N_{\rm crit}$~\cite{Boissonneault09}, where $N_{\rm crit} = [(\omega_0-\omega_{\rm c})/(2g)]^2 \approx 30$ for our experimental values.

At the transition and further in the classical regime, the two subsystems become strongly hybridized. In this case $[\hat{H},\hat{\sigma}_{\rm z}] \neq 0$ and, therefore, the qubit population is no longer conserved. We have observed this effect by performing relaxation measurements: under continuous constant driving at $\omega_{\rm d}=4.366$ GHz (see Fig.~\ref{fig:schema}(d)), we excite the system with a long pulse at the ac-Stark shifted qubit frequency and we monitor the response of the cavity after this pulse ends.  We extract the relaxation time from fitting the relaxation traces with an exponential, see Fig.~\ref{fig:expBSshift}(b). The results show an increase in the relaxation time as the power is ramped up to above a cavity photon number $N_{c}\approx 1$, coinciding with the onset of the quantum-to-classical transition. In the same power range the qubit frequency decreases according to the standard quantum theory of ac-Stark shifted transitions~\cite{Boissonneault09,Boissonneault12a,Boissonneault12b} which also includes effects caused by the higher transmon energy levels. At about the same number of photons, discrepancies with respect to this model start to appear. A classical theory~\cite{supplement,Irwin,Irwin2} which treats the transmon as an anharmonic oscillator provides a good fitting with $T_1$ up to the power level when the cavity frequency can no longer be taken as constant (about $N_c \approx 2$, after which the measured $T_1$ drops again). 

In Fig.~\ref{fig:expBSshift}(c), we demonstrate that the fluctuations $\langle \delta N^2\rangle = \langle N^2\rangle - \langle N \rangle^2$ of the transmon occupation number change rather abruptly when the drive power is swept over the transition regime. Thus, distinctly from the transition predicted by Carmichael~\cite{Carmichael15}, also the transmon becomes a multilevel classical object. In the laboratory frame, we characterize the quantum-to-classical transition by defining a dimensionless order parameter $\Xi=|\langle (\hat{a}+\hat{a}^{\dag})\hat{\Pi}_{\rm x}\rangle|$, where $\hat{\Pi}_{\rm x}=\hat{c}+\hat{c}^{\dag}$ is the analog of $\hat{\sigma}_{\rm x}$ and $\hat{c}$ is the annihilation operator for the multilevel transmon~\cite{BishopPhD}. Note that this order parameter includes the correlations between the cavity and the transmon. At low powers the order parameter saturates to a nearly-zero value given by $\Xi_{\rm ground}\approx 2g/(\omega_0+\omega_{\rm c}) \approx 10^{-2}$, indicating the quantum regime. As the power increases, the order parameter starts to increase at $N_{\rm c}\approx 1$ and the fluctuations become large as well. At high enough power (approximately $N_{\rm c}=10$), the system enters the classical phase, where a large number of states participate in the dynamics. Interestingly, the BS effect is reproduced by the order parameter which is seen as the nonmonotonic deviation between the RWA and non-RWA order parameters.

\begin{figure}[th!]
\centering
\includegraphics[width=1.0\linewidth]{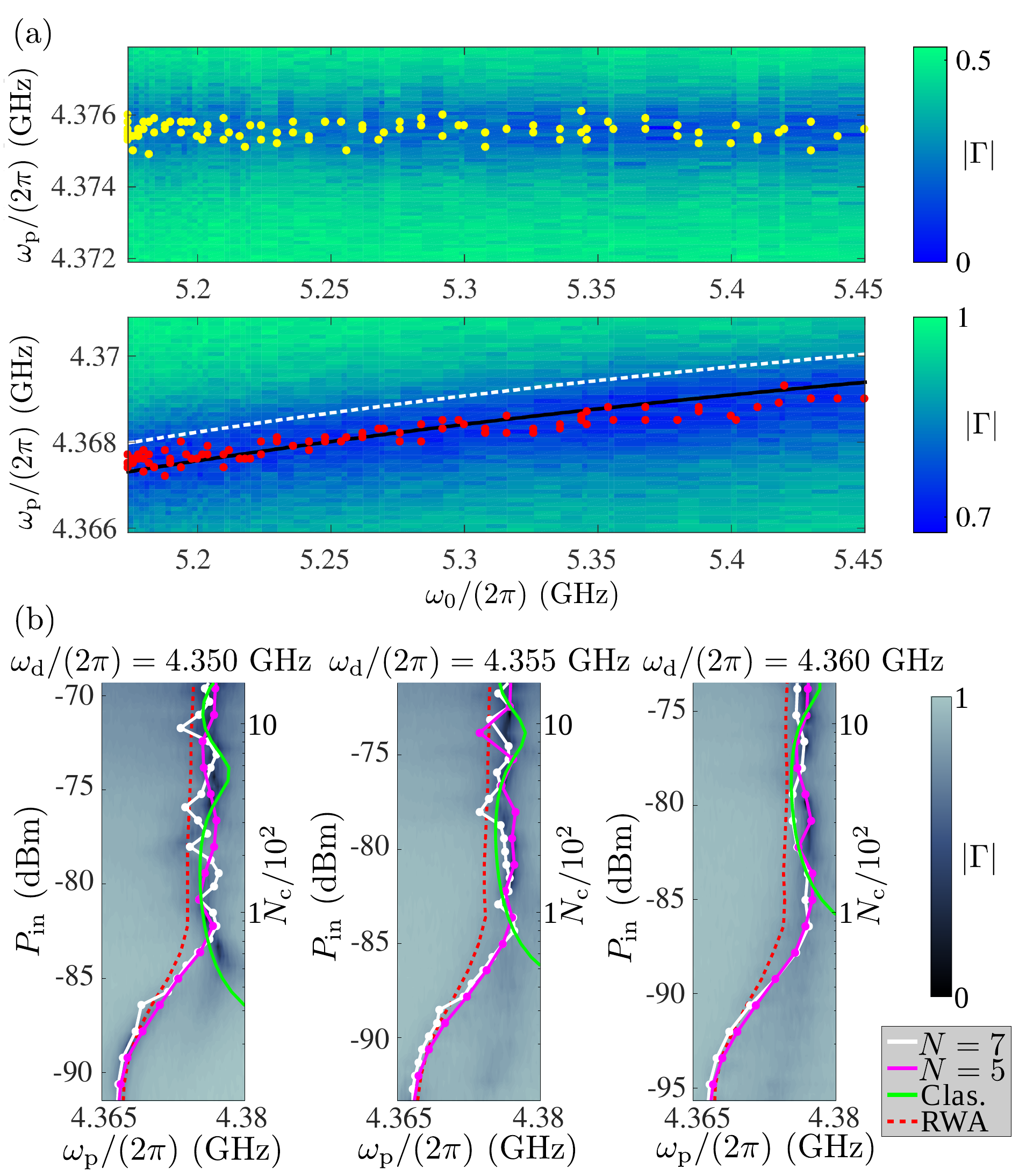}
\caption{The Bloch--Siegert effect in the quantum and classical regimes. (a) Measured reflection coefficient $|\Gamma|$ as a function of $\omega_0$ and $\omega_{\rm p}$, taken at high power (upper panel) and low power (lower panel). The strongest resonance minima extracted from the data are marked with dots. The lower panel demonstrates the vacuum BS shift of $\chi_{\rm BS} = 0.7$ MHz, displayed as the difference between the calculated cavity frequency with only the analytic vacuum-Stark shift ($\omega_{\rm c}-\chi_0$, dashed white line) and with the analytic vacuum-BS shift included ($\omega_{\rm c}-\chi_0-\chi_{\rm BS}$, solid black line). (b) Bloch--Siegert shift as a function of driving power (corresponding to $N_{\rm c}=10\ldots 1650$) in the classical regime for $\omega_{\rm d}/(2\pi) = 4.350, \ 4.355, \ \textrm{and} \ 4.360$ GHz. We show the experimental contour for the reflection coefficient and compare the resonance locations with the classical-approximation formula for $\omega_+$ in Eq.~(\ref{eq:class}) and with the numerical results with $N=5,7$ ($N=7$ is calculated also in the RWA).}\label{fig:Boxes}
\end{figure}

In Fig.~\ref{fig:Boxes}(a), we further demonstrate the essential difference between the quantum and classical regimes and the role of the Bloch--Siegert shift by analyzing the dependence of the response on the qubit frequency at the two extremes of the drive power. As expected, at very high powers the resonance is located at the bare cavity frequency, irrespective of the qubit frequency. At low drive powers, the resonance location decreases with the qubit frequency, which is characteristic to a vacuum-Stark and BS shifted cavity. The inclusion of the vacuum-BS shift clearly gives a better fit to the resonance data.  We have also studied in more detail the BS effect in the transition region. In addition to numerical results, we can  also employ the classical model to calculate the normal modes 
\begin{equation}\label{eq:class}
\omega_{\pm} = \sqrt{\frac{\Omega^2+\omega_{\rm c}^2}{2}
\pm\frac{1}{2}\sqrt{\left[\Omega^2-\omega_{\rm c}^2\right]^2+16g^2\Omega^2\omega_{\rm c}/\omega_0} },
\end{equation}
where $\Omega^2 = J_0(\sqrt{P_{\rm in}/P_0})\omega_{0}^2$, $P_{\rm in}$ is the power fed into the system, and $P_0=1.0$ pW. These are obtained by treating the Josephson junction as a classical inductance and by averaging over the drive period~\cite{supplement,Saira16}. 
In Fig.~\ref{fig:Boxes}(b) we present the results of experiments
at three different drive frequencies. Because the transition occurs at a fixed value $N_{\rm c}\sim 1$ for all drive frequencies~\cite{supplement}, we show the responses in the same interval of cavity quanta, ranging from $N_{\rm c}=10\ldots 1650$. One notices that in all three measurements the BS effect is present. There is very good agreement between the experimental data and the numerics, and also in the high-power region the classical mode $\omega_+$ provides a good match to the data.

To conclude, we have studied a driven-dissipative quantum-to-classical transition in a circuit-QED setup consisting of a transmon coupled to a cavity. The response of the system to a weak probe reveals a clear signature of the Bloch--Siegert shift, which is nonmonotonous in the drive power. 
Our experiment paves the way toward future experiments of quantum simulations of dissipative phase transitions in one-dimensional circuit-QED lattices.

\begin{acknowledgments}
Discussions with D. Angelakis, E. Thuneberg, M. Silveri, and S. Laine on theory, and M. Sillanp\"a\"a and J. Pirkkalainen on sample design and fabrication are gratefully acknowledged. We are grateful for the financial support from the Academy of Finland (projects No. 263457 and No. 135135), the Magnus Ehrnrooth Foundation, the Finnish Cultural Foundation, Väisälä Foundation, FQXi, Centre of Quantum Engineering at Aalto University (project QMET), and the Centres of Excellence LTQ (project No. 250280) and COMP (projects No. 251748 and No. 284621). This work used the cryogenic facilities of the Low Temperature Laboratory at OtaNano/Aalto University.
\end{acknowledgments}



\pagebreak
\widetext
\begin{center}
\textbf{\large Observation of the Bloch--Siegert shift in a driven quantum-to-classical transition: supplementary information}
\end{center}
\setcounter{equation}{0}
\setcounter{figure}{0}
\setcounter{table}{0}
\setcounter{page}{1}
\makeatletter
\renewcommand{\theequation}{S\arabic{equation}}
\renewcommand{\thefigure}{S\arabic{figure}}
\renewcommand{\bibnumfmt}[1]{[S#1]}
\renewcommand{\citenumfont}[1]{S#1}




\newcommand{\matr}[1]{\mathsf{#1}}
\newcommand{\vekt}[1]{\mathbf{#1}}
\newcommand{\unitvekt}[1]{\mathbf{\hat{#1}}}
\newcommand{\uvekt}[1]{\mathbf{\hat{#1}}}
\newcommand{\erik}[1]{\mathrm{#1}}
\newcommand{\ee}{\mathrm{e}}
\newcommand{\ii}{\mathrm{i}}
\newcommand{\D}[1]{\,\text{d}#1\,}
\newcommand{\Dn}[2]{\,\text{d}^{#1}#2\,}
\newcommand{\bra}[1]{\mathinner{\langle{#1}|}}
\newcommand{\ket}[1]{\mathinner{|{#1}\rangle}}
\newcommand{\inner}[3]{\left< #1\left|#2\right|#3\right >}
\newcommand{\braket}[2]{\langle #1 | #2 \rangle}
\newcommand{\Bra}[1]{\left<#1\right|}
\newcommand{\Ket}[1]{\left|#1\right >}
\newcommand{\dt}[1]{\frac{\text{d}#1}{\text{d}t}}
\newcommand{\ot}[1]{\hat{#1}}
\newcommand{\rt}[1]{{\rm #1}}
\newcommand{\upa}{\uparrow}
\newcommand{\downa}{\downarrow}
\newcommand{\upup}{\uparrow\!\uparrow}
\newcommand{\updown}{\upa\!\downa}
\newcommand{\downup}{\downa\!\upa}
\newcommand{\downdown}{\downa\!\downa}
\newcommand{\expe}[1]{\left\langle #1 \right\rangle}
\newcommand{\expes}[1]{\langle #1 \rangle}
\newcommand{\Tr}[1]{\textrm{Tr}\left( #1\right)}





\section*{Experimental methods}

\begin{figure}[h!]
  \centering
\includegraphics[width=0.8\linewidth]{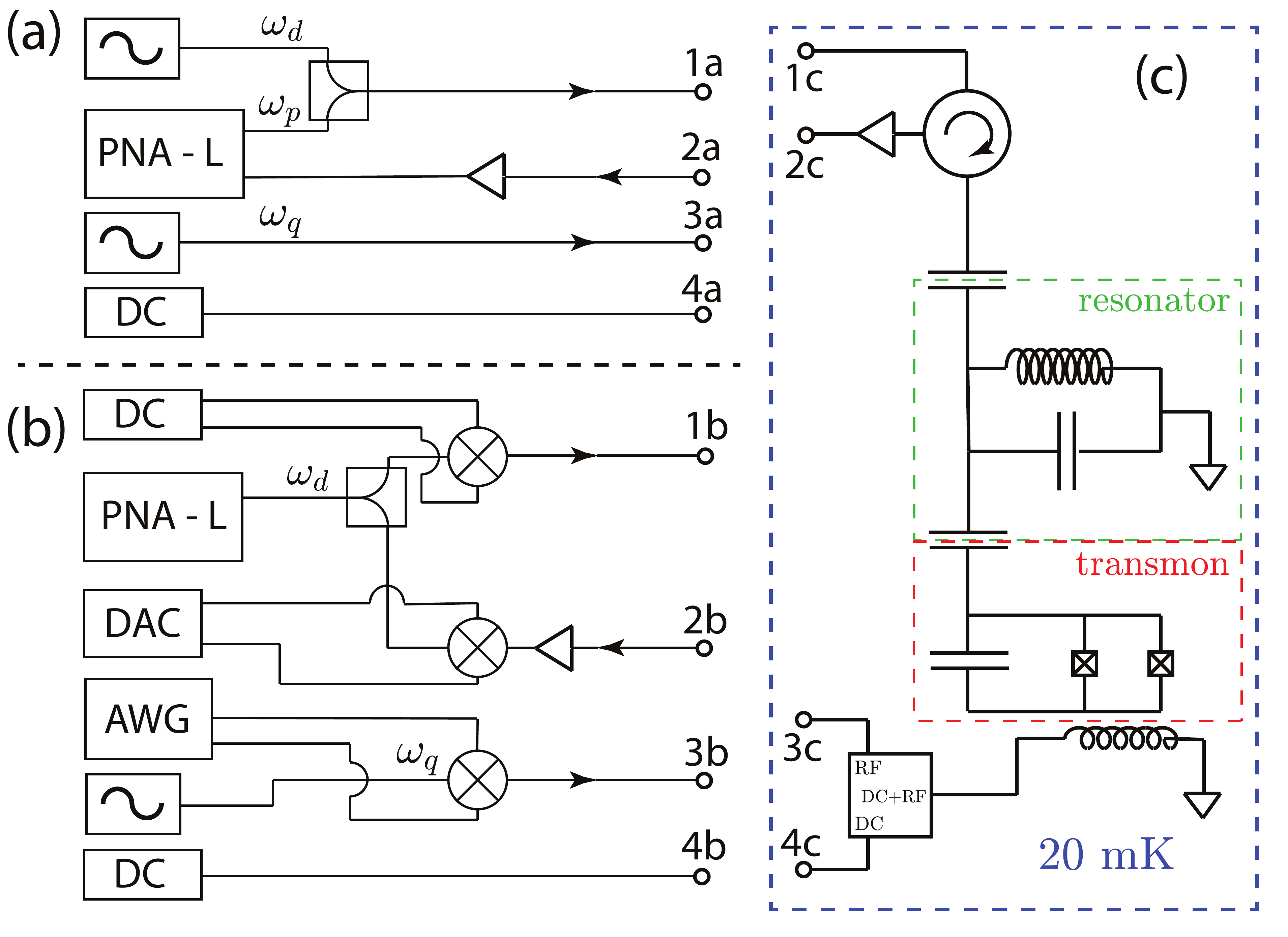}
\caption{Simplified circuit schematic of the measurement setup. We show (a) the room-temperature control electronics for the reflection measurement and; (b) the relaxation time measurement. (c) Low-temperature circuit. Here $\omega_{\rm q}$ refers generically to the frequency of the microwave tone applied to the qubit (port 3c of the low temperature setup), while $\omega_{\rm d}$ and $\omega_{\rm p}$ are the drive and probe frequencies.}\label{fig:mschema}
\end{figure}

In Fig.~\ref{fig:mschema}, we present the electronic measurement set	up at the room temperature used for the reflection spectrometry, and for the measurements of the relaxation time. We also show 
show the electrical circuit at low temperature (including the transmon). For the spectroscopy measurements [Fig.~\ref{fig:mschema}(a)], the cavity probe signal with frequency $\omega_{\rm p}$ is generated by an Agilent N5230C PNA-L network analyzer and the cavity drive with frequency $\omega_{\rm d}$ is provided by an Agilent E8257D analog signal generator. These are combined 
in a Mini-Circuits ZFSC-2-10G power splitter/combiner and sent to the cavity input line of the dilution refrigerator. The tone $\omega_{\rm q}$ is used for qubit characterization by two-tone spectroscopy with the drive tone turned off.

In the case of relaxation time measurements shown in Fig.~\ref{fig:mschema}(b), we use the signal provided by the PNA-L network analyzer to drive the resonator and at the same time to provide a frequency reference for the mixer used for down-conversion. To create the microwave pulse applied to the qubit, we use an arbitrary waveform generator Tektronix AWG5014B. The pulse-shape signal created by this device is up-converted to the qubit frequency using a mixer and the local oscillator (LO) provided by the Agilent E8257D analog signal generator. As shown in Fig.~1(d) in the main text, we produce in this way a 2 $\mu\textrm{s}$-long microwave excitation pulse which is applied to the transmon gate line. The frequency $\omega_{\rm q}$ of this pulse is kept in resonance with the ac-Stark shifted transition frequency between the ground and first excited states of the transmon, which was measured separately. The incoming signal from the resonator is down-converted using an IQ mixer with the LO at $\omega_{\rm d}$, and then digitalized by a fast data aquisition card Agilent Acqiris U1084A. 
For each value of the resonator drive power, the transmon state preparation and resonator response measurement are repeated $13\times 10^6$ times and averaged. This scheme allows us to monitor in time domain the variation of the reflected signal over several microseconds. After the trailing edge of the pulse, we record the relaxation traces of the system, and we extract the $T_1$ time by fitting with an exponential. In all these traces, small oscillations exist immediately after the trailing edge of the pulse, leading to the fitting errors shown in Fig.~3(b) in the main text. This procedure is repeated at 95 drive power values.

For tuning the qubit transition frequency, in both cases depicted in Figs.~\ref{fig:mschema}(a) and~\ref{fig:mschema}(b) the dc flux bias is provided by an Agilent 33500B waveform generator. This is combined with the qubit drive using a bias-tee and applied to the transmon gate, see Fig.~\ref{fig:mschema}(c). The signal from
the cavity output line of the dilution refrigerator is separated from the input signal using a cryogenic circulator, and then
amplified by cryogenic and room-temperature amplifiers. DC blocks (not shown) are used to break all possible ground loops between the microwave instruments (analog signal
generator and network analyzer) and their corresponding lines. Attenuators (not shown) are added between every connection to reduce cable resonances. All instruments are synchronized with a SRS FS725 Rubidium frequency standard (not shown). Measurement
controlling and data processing are done by MATLAB running on a measurement
computer. Communications between the measurement computer and the instruments
are realized through IEEE-488 GPIB buses.



\subsection{Power attenuation}

In our experiments, we drive the system strongly with a signal created by the signal generator at room temperature. We can control the signal power at the generator, but to get the power $P_{\rm in}$ at the input of the cavity we need to make an estimation of the attenuation in the cavity input line.

In both spectroscopy and relaxation experiments, we use the same line to connect the ports 1a to 1c and 1b to 1c (see Fig.~\ref{fig:mschema}). There are several attenuators placed in this line at the different temperature levels of the refrigerator, providing a total of $68\ \textrm{dB}$ attenuation. Then for the semi-rigid cables of this rf line we have $9\ \textrm{dB}$ of additional attenuation (calculation includes the temperature dependence of attenuation). Further on, as can be seen in Fig.~\ref{fig:mschema}(c), a cryogenic circulator was used at the mixing chamber plate to separate the input signal from the signal reflected by the cavity. The circulator had an insertion loss of $0.35\ \textrm{dB}$. Finally, the connection between the PCB of the sample holder and the on-chip cavity input/output line was done with two aluminum wire bonds $25\ \mu\textrm{m}$ in diameter and roughly $2\ \textrm{mm}$ in length. These wire bonds cause impedance mismatch and as a result some reflection occurs at each end of the wire bond. The power which reaches the on-chip resonator cavity will be reduced due to this reflection, and the reduction in power can be expressed as an equivalent attenuation of
$$
10\cdot \log_{10}\left(\left[1 - \frac{|Z_w - 50\ \Omega|}{Z_w + 50\ \Omega}\right]^2\right)\ [\textrm{dB}],
$$
where $Z_w \simeq \sqrt{\frac{L_w}{C_w}}$ is the wave impedance of the wire-bond connection, $L_w$ and $C_w$ are the inductance and the capacitance of the wire bond connection. The estimate of the inductance of each wire is $1.9\ \textrm{nH}$, and the capacitance of the wire is of the order of $25\ \textrm{fF}$. These values give $Z_w = 137\ \Omega$ for the two wire bonds, resulting in $10.8\ \textrm{dB}$ for the 
attenuation of the wire bond connection.

As a result, our estimation of the total attenuation from the ports 1a and 1b to the input of the cavity is 88.5 dB, i.e. $P_{\rm in, a} \approx P_{1{\rm a}}-88.5$ dBm in the spectroscopy experiment and similarly $P_{\rm in, b} \approx P_{1{\rm b}}-88.5$ dBm in the relaxation experiment. In the spectroscopy experiment, the power $P_{1{\rm a}}$ is obtained from the readout of the analog signal generator, from which we substract $3.5\ \textrm{dB}$, representing the insertion loss of the power combiner. 
In the case of relaxation measurements, the power $P_{1{\rm b}}$ is obtained experimentally from mixer calibration (measuring the power at 1b for different DC voltages applied to the I port in the linear regime of the mixer).
The estimates above are remarkably consistent with results obtained from fitting the spectroscopy data in Fig. 3(a) and the ac Stark-shifted qubit frequency [inset of Fig.~3(b)] in the main text, which correspond to an attenuation of $\sim 85.5$ dB in the line 1a-1c (1b-1c). In this sense, any one of these plots can be regarded as a precise power calibration for the others.


\section{Transmon characterization}
The classical Hamiltonian function for our driven transmon-cavity system can be written into the form~\cite{S_Koch07}
\begin{equation}
H=\frac{\Phi_{\rm r}^2}{2L_{\rm r}}+\frac{Q_{\rm r}^2}{2C_{\rm r}}+\frac{Q_{\rm J}^2}{2C_{\Sigma}}-E_{\rm J}\left(\Phi\right)\cos\left(\frac{2\pi}{\Phi_0}\Phi_{\rm J}\right)+C_{\rm g} \frac{Q_{\rm r}Q_{\rm J}}{C_{\rm r}C_{\Sigma}}+\frac{C_{\rm c}}{C_{\rm r}} Q_{\rm r} V_{\rm in}(t),
\end{equation}
where the Josephson energy $E_{\rm J}(\Phi)$ is controlled by the external magnetic flux $\Phi$ applied through the SQUID loop of the transmon. In the above Hamiltonian, $V_{\rm in}(t)=V_{\rm in} \cos(\omega_{\rm d} t)$ is the effective drive caused by the transmission line and the voltage amplitude can be written in terms of the input power at the sample as $V_\text{in} =\sqrt{2Z_0 P_\text{in}}$. Also, we denote by $\Phi_{\rm r}$, $\Phi_{\rm J}$ and $Q_{\rm r}$, $Q_{\rm J}$ the magnetic fluxes and the electric charges of the resonator and the transmon qubit, respectively~\cite{S_Koch07}. The parameters $C_{\rm r}$ and $L_{\rm r}$ are the capacitance and inductance of the resonator, respectively, $C_{\Sigma}$ is the capacitance of the transmon, $\Phi_0 = h/(2e)$ is the magnetic flux quantum, $C_{\rm g}$ is the gate capacitance of the transmon, and $C_{\rm c}$ is the coupling capacitance between the transmission line and the resonator.

The canonical quantization procedure results in the Hamiltonian (scaled with $\hbar$)
\begin{equation}
\label{eq:S_transHam}
\hat{H}= \omega_{\rm c} \hat{a}^{\dag}\hat{a} + \sum_{n=0}^{N-1} \Omega_{n} \hat{\Pi}_{nn} + (\hat{a}^{\dag}+\hat{a})\sum_{n,m=0}^{N-1} g_{nm}\hat{\Pi}_{nm}+A\cos(\omega_{\rm d} t)(\hat{a}^{\dag}+\hat{a}),
\end{equation}
where we have diagonalized the transmon part of the system and denoted with $\omega_{\rm c}=1/\sqrt{L_{\rm r}C_{\rm r}}$ and $\hat{a}$ the natural angular frequency and the annihilation operator of the resonator, respectively. Also, we denote the eigenenergies and the corresponding eigenstates of the transmon with $\Omega_{n}$ and $|n\rangle$, respectively, where $n=0,1,\ldots N-1$ with $N$ being the number of transmon eigenstates included into the model. The ladder operators for the transmon are defined as $\hat\Pi_{nm}\equiv |n\rangle\langle m|$. With these notations, the first transition frequency of the transmon is $\omega_0/(2\pi) =(\Omega_1-\Omega_0)/(2\pi)$.
The quantized charge of the resonator can be written as $\hat{Q}_{\rm r} =C_{\rm r} V_{\rm rms}^0(\hat{a}^{\dag}+\hat{a})$, which allows us to define the coupling coefficient $g_{nm} = \frac{2eC_{\rm g}}{C_{\Sigma}}V_{\rm zpf} \langle n|\hat{Q}_{\rm J}/(2e)|m\rangle$. We have defined the zero-point fluctuations of the resonator voltage as
$V_{\rm zpf}=\sqrt{\frac{\hbar \omega_{\rm c}}{2C_{\rm r}}}$. The calibration between the drive amplitude $A$ and the power $P_{\rm in}$ at the input of the cavity can be written as
\begin{equation}
A = \frac{C_{\rm c}}{\hbar}\sqrt{\frac{\hbar \omega_{\rm c} Z_0 P_\text{in}}{C_{\rm r}}}.
\end{equation}
In our system, the input capacitance $C_{\rm c} \simeq 12$ fF, the natural cavity frequency $\omega_{\rm c}/(2\pi) \simeq 4.376$ GHz, the impedance of the transmission line $Z_0 \simeq 50 \ \Omega$, and the resonator capacitance $C_{\rm r} \simeq 1.24$ pF.

Numerically, the transmon energies and eigenstates can be obtained from the Mathieu equation~\cite{S_Koch07} as $\hbar\Omega_n = E_{\rm C} \mathcal{M}_{\rm A}(2(n_{\rm g}+k(n,n_{\rm g})),-E_{\rm J}/(2E_{\rm C}))$, where $\mathcal{M}_{\rm A}(r,q)$ is Mathieu's characteristic value, and $k(n,n_{\rm g}) = \sum_{l=\pm 1} [\text{int}(2n_{\rm g} +l/2)\text{mod}\,2][\text{int}(n_{\rm g})+l(-1)^n((n+1)\text{div}\,2)]$. Here int($x$) rounds to a nearest integer, $a$ mod $b$ is the typical modulo operator and $a$ div $b$ is the quotient of $a$ and $b$. The coupling coefficients can be solved in terms of the Mathieu functions as $g_{nm} = \lambda\int_0^{2\pi} d\varphi_{\rm J}\psi^\ast(n,\varphi_{\rm J})(-i\frac{\D}{\D{\varphi_{\rm J}}})\psi(m,\varphi_{\rm J})$, where $\lambda=\frac{2eC_{\rm g}V_{\rm zpf}}{C_{\Sigma}}$, $\varphi_{\rm J}=2\pi\Phi_{\rm J}/\Phi_0$ is the superconducting phase difference across the transmon, and $\psi(n,\varphi_{\rm J}) = \frac{e^{in_{\rm g}\varphi_{\rm J}}}{\sqrt{2\pi}} [\mathcal{M}_{\rm C}(\frac{\hbar\Omega_n}{E_{\rm C}},-\frac{E_{\rm J}}{2E_{\rm C}},\frac{\varphi_{\rm J}}{2}) +i(-1)^{k+1}\mathcal{M}_{\rm S}(\frac{\hbar\Omega_n}{E_{\rm C}},-\frac{E_{\rm J}}{2E_{\rm C}},\frac{\varphi_{\rm J}}{2})]$, where $\mathcal{M}_{\rm C}(r,q,\theta)$ and $\mathcal{M}_{\rm S}(r,q,\theta)$ are the Mathieu cosine and sine functions, respectively \cite{S_CottetPhD}. In our measurement setup, the transmon is biased using a magnetic field such that $E_{\rm J}(\Phi)/E_{\rm C} \simeq 28$ which implies that coupling coefficients $g_{nm}$ between nonadjacent transmon states are negligible. To simplify the notations, we denote $E_{\rm J}\equiv E_{\rm J}(\Phi)$ from now on. 
Also, our parameter values $C_{\Sigma} \simeq 50$ fF, and $C_{\rm g} \simeq 20$ fF indicate that $\lambda/h = 90$ MHz resulting in $g_{01}/(2\pi) = 80$ MHz.

%

\subsection{Low-power spectroscopy}

\begin{figure}[h!]
  \centering
\includegraphics[width=0.8\linewidth]{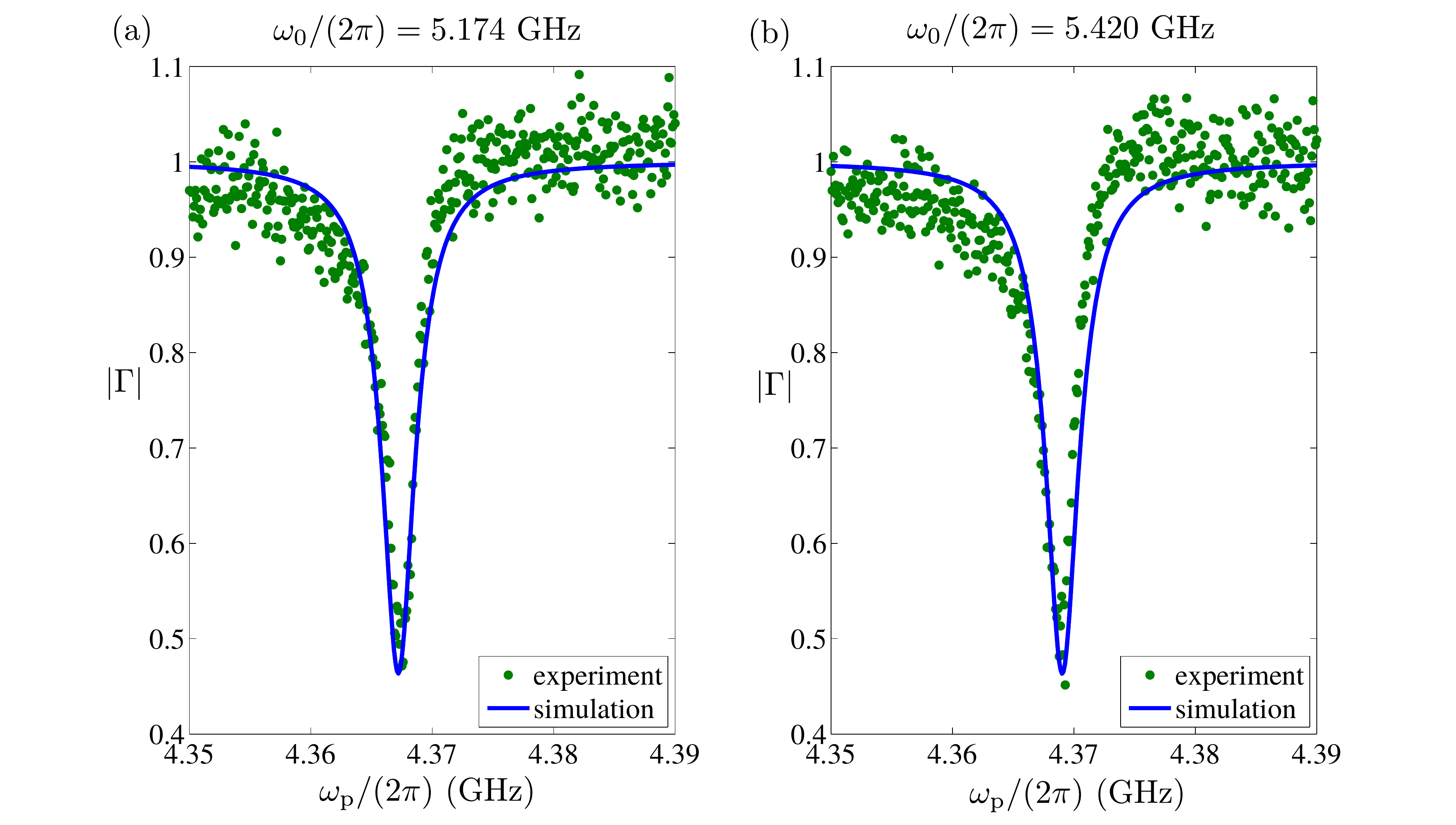}
\caption{Low-power reflection spectrum. We plot the spectral line of the cavity for two flux values, one corresponding to the minimum qubit frequency and the other one to the highest qubit frequency at which the qubit line is still observable in the two-tone spectroscopy.}\label{fig:lowpower}
\end{figure}

\begin{figure}[h!]
  \centering
\includegraphics[width=0.8\linewidth]{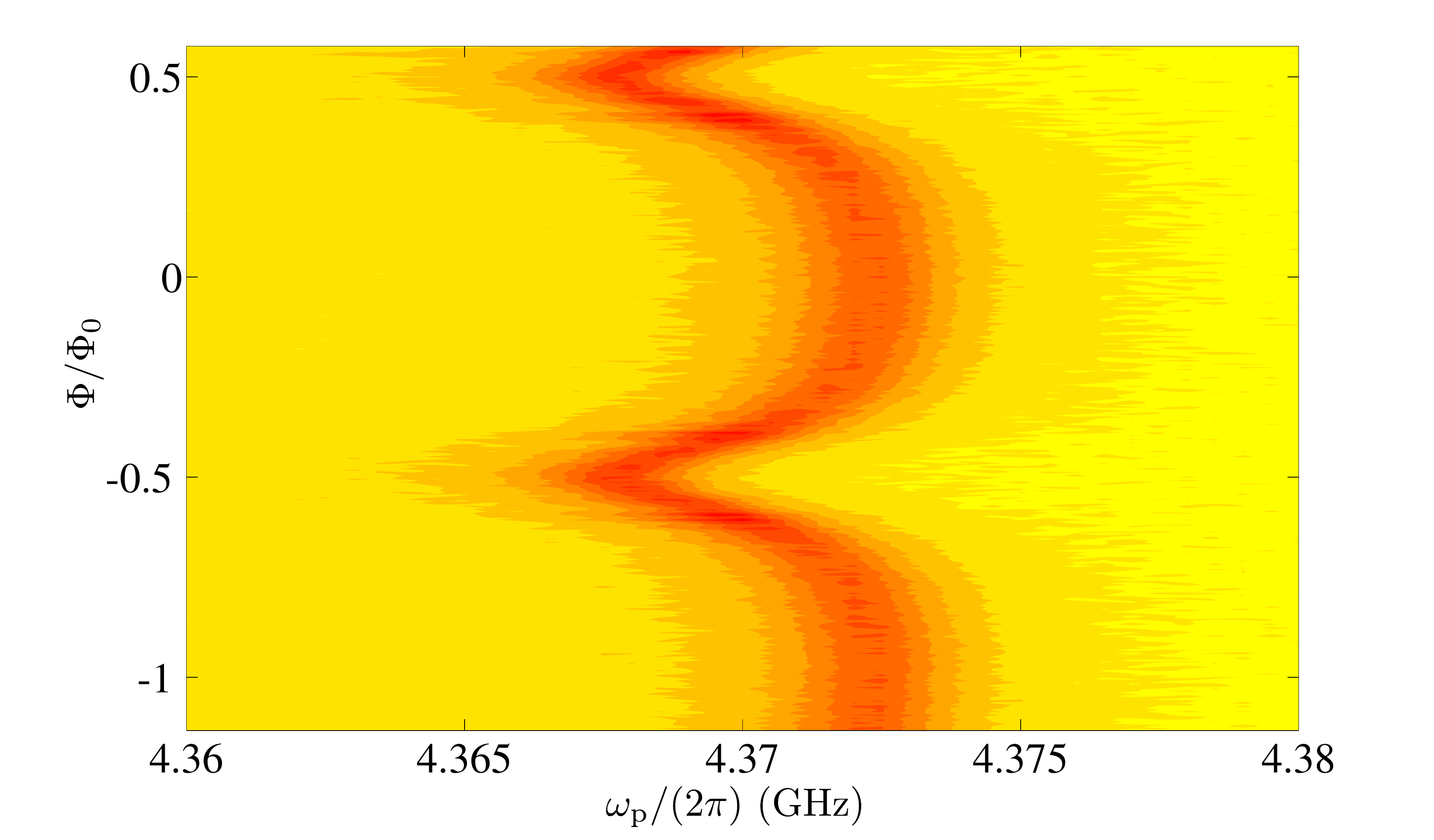}
\caption{Spectral line of the cavity as a function of the external flux $\Phi$.}\label{fig:ResonatorLine}
\end{figure}

We can estimate our circuit parameters by measuring the reflection coefficient (in the absence of the strong resonator drive tone $\omega_{\rm d}$) at the transition frequencies $\omega_0/(2\pi) =5.174$ GHz and $\omega_0/(2\pi)=5.42$ GHz between the ground state and the first excited state (see the definition of $\omega_0$ above). The data is presented in Fig.~\ref{fig:lowpower}. Fitting the simulation with the data, we obtain the estimates $\kappa/(2\pi)= 4$ MHz and $g_{01}/(2\pi)= 80$ MHz.
In Fig.~\ref{fig:ResonatorLine} we show the $\Phi_0$-periodic flux dependence of the resonator response, probed with the tone $\omega_{\rm p}$.

\subsection{Displacement transformation}

Typically, when the driving is strong, it is beneficial to make a displacement transformation for the cavity with
\begin{equation}\label{eq:S_disp}
\hat{D}(\alpha) = e^{\alpha \hat{a}^{\dag}-\alpha^* \hat{a}},
\end{equation}
where $\alpha$ is generally complex and time-dependent. In order to retain the form of the Schr\"odinger equation, the Hamiltonian in Eq.~(\ref{eq:S_transHam}) transforms as $\hat{H}\rightarrow \hat{D}^{\dag}\hat{H}\hat{D} + i\dot{\hat{D}}^{\dag}\hat{D}$. After the transformation, the Hamiltonian can be written as
\begin{eqnarray}
\hat{H}&=& \omega_{\rm c} \hat{a}^{\dag} \hat{a} + \sum_{n=0}^{N-1} \Omega_{n} \hat{\Pi}_{nn} + (\hat{a}^{\dag}+\hat{a})\sum_{n,m=0}^{N-1} g_{nm}\hat{\Pi}_{nm}  + (\alpha^{*}+\alpha)\sum_{n,m=0}^{N-1} g_{nm}\hat{\Pi}_{nm}\\
&& + \omega_{\rm c}(\alpha^* \hat{a} + \alpha\hat{a}^{\dag}) + \frac{A}{2}(\hat{a}^{\dag}e^{-i\omega_{\rm d} t}+\hat{a}e^{i\omega_{\rm d} t}) + i\left(\dot{\alpha}^*\hat{a}-\dot{\alpha} \hat{a}^{\dag}\right),
\end{eqnarray}
where we have also made a rotating-wave approximation for the drive. By making the displacement transformation also for the master equation of the density operator, one obtains the master equation in the displaced frame:
\begin{equation}
\begin{split}
\frac{d\hat{\rho}}{dt}=& -i [\hat{H}_{\rm eff},\hat{\rho}] + \kappa \mathcal{L}[\hat{a}]\hat{\rho} + \gamma \mathcal{L}[\hat{\sigma}_-]\hat{\rho} + \frac{\gamma_{\phi}}{2} \mathcal{L}[\hat{\sigma}_{\rm z}]\hat{\rho}\\
&-i[\hat{a},\hat{\rho}]\left(i\dot{\alpha}^*+\omega_{\rm c}\alpha^*+\frac{A}{2}e^{i\omega_{\rm d} t} +i\frac{\kappa}{2}\alpha^*\right)\\
&-i[\hat{a}^{\dag},\hat{\rho}]\left(-i\dot{\alpha}+\omega_{\rm c}\alpha+\frac{A}{2}e^{-i\omega_{\rm d} t} -i\frac{\kappa}{2}\alpha\right)\\
=& - i [\hat{H}_{\rm eff},\hat{\rho}] + \kappa \mathcal{L}[\hat{a}]\hat{\rho} + \gamma \mathcal{L}[\hat{\sigma}_-]\hat{\rho} + \frac{\gamma_{\phi}}{2} \mathcal{L}[\hat{\sigma}_{\rm z}]\hat{\rho},
\end{split}
\end{equation}
where the Lindblad superoperator $\mathcal{L}[\hat{A}]\hat{\rho} = \frac12 (\hat{A}\hat{\rho} \hat{A}^{\dag} - \hat{A}^{\dag}\hat{A}\hat{\rho}-\hat{\rho} \hat{A}^{\dag}\hat{A})$ and
\begin{equation}
\hat{H}_{\rm eff} \equiv \omega_{\rm c} \hat{a}^{\dag} \hat{a} + \sum_{n=0}^{N-1} \Omega_{n} \hat{\Pi}_{nn} + (\hat{a}^{\dag}+\hat{a})\sum_{n,m=0}^{N-1} g_{nm}\hat{\Pi}_{nm}  + (\alpha^{*}+\alpha)\sum_{n,m=0}^{N-1} g_{nm}\hat{\Pi}_{nm} .
\end{equation}
The last equality holds if we choose that $\alpha$ obeys the classical equation of motion of the driven and damped harmonic oscillator:
\begin{equation}
\dot{\alpha} = -i\omega_{\rm c} \alpha -i \frac{A}{2}e^{-i\omega_{\rm d} t} - \frac{\kappa}{2}\alpha.
\end{equation}
Notice that this choice can always be made since we did not define $\alpha$ in Eq.~(\ref{eq:S_disp}). The solution of the above equation in the steady state is
\begin{equation}\label{eq:S_ass}
\alpha_{\rm ss} = \frac{-Ae^{-i\omega_{\rm d} t}}{2(\delta_{\rm c} - i\frac{\kappa}{2})}=\frac{Ae^{-i\omega_{\rm d} t}}{2\sqrt{\delta_{\rm c}^2+\kappa^2/4}},
\end{equation}
where 
we have denoted the detuning between the cavity and the drive with $\delta_{\rm c} = \omega_{\rm c}-\omega_{\rm d}$ and assumed, for simplicity, that the drive has a constant phase shift $\theta = \arctan(\kappa/(2\delta_{\rm c}))$. The total number of photons in the cavity is $N_{\rm c} = |\alpha_\text{ss}|^2 +\langle \hat{a}^\dagger\hat{a} \rangle$. Also, the effective Hamiltonian in the displaced frame can be written as
\begin{equation}\label{eq:S_Hameffmulti}
\hat{H}_{\rm eff} = \omega_{\rm c} \hat{a}^{\dag} \hat{a} + \sum_{n=0}^{N-1} \Omega_{n} \hat{\Pi}_{nn} + (\hat{a}^{\dag}+\hat{a})\sum_{n,m=0}^{N-1} g_{nm}\hat{\Pi}_{nm}  + \sum_{n,m=0}^{N-1} G_{nm}\cos(\omega_{\rm d} t)\hat{\Pi}_{nm},
\end{equation}
where $G_{nm} = 2g_{nm}|\alpha_{\rm ss}|$. This is the Hamiltonian used in the numerical calculations. By truncating the transmon eigenbasis to the two lowest states, we obtain
\begin{equation}
\label{eq:S_Hameff}
\hat{H}_{\rm eff} = \omega_{\rm c} \hat{a}^{\dag}\hat{a} + \frac{\omega_0}{2}\hat{\sigma}_{\rm z} + g(\hat{a}^{\dag}+\hat{a})\hat{\sigma}_{\rm x} + \frac{G}{2}\left[e^{i\omega_{\rm d} t}+e^{-i\omega_{\rm d} t}\right]  \hat{\sigma}_{\rm x},
\end{equation}
where we denote $g=g_{01}$, $\omega_0 = \Omega_1-\Omega_0$ and $G\equiv G_{01}=gA/\sqrt{\delta_{\rm c}^2+\kappa^2/4}$. The last term clearly excites the qubit when the driving is strong. In terms of the parameters in our experiment, i.e. $\omega_0/(2\pi) = 5.16$ GHz; $g/(2\pi) = 80$ MHz and $N_{\rm c} \geq 10^3$, we obtain
\begin{equation}
G/(2\pi)\geq 5 \textrm{ GHz}.
\end{equation}
Thus, we see that the qubit can become highly excited at the high-power ends of our spectra. 

We then make the Schrieffer--Wolff transformation for Hamiltonian in Eq.~(\ref{eq:S_Hameff}) with $\hat{U}=\exp[-\frac{g}{\omega_0-\omega_{\rm c}}(\hat{a}\hat{\sigma}_+-\hat{a}^{\dag}\hat{\sigma}_-)]$, where $\hat{\sigma}_-$ is the annihilation operator of the qubit. By retaining terms up to second order in $g$ we obtain Eq.~(2) in the main text.

\subsection{Counter-rotating hybridized RWA}

Instead of applying the RWA for Eq.~(\ref{eq:S_Hameff}), we make a transformation into a nonuniformly rotating frame with~\cite{S_Lu12, S_Yan15}
\begin{equation}
\hat{U}=\exp(-\hat{S})\equiv \exp\left(-i \frac{G}{\omega_{\rm d}}\xi \sin(\omega_{\rm d} t) \hat{\sigma}_{\rm x}\right),
\end{equation}
where $\xi$ is a free parameter to be determined later. As a consequence, $\hat{H}\rightarrow \hat{U}^{\dag} \hat{H} \hat{U} + i \dot{\hat{U}}^{\dag} \hat{U}$ and the transformed Hamiltonian can be written as
\begin{equation}
\hat{H}=\omega_{\rm c} \hat{a}^{\dag}\hat{a} + \frac{\omega_0}{2}\left[\cos\left(\frac{2G}{\omega_{\rm d}}\xi\sin(\omega_{\rm d} t)\right)\hat{\sigma}_{\rm z} + \sin\left(\frac{2G}{\omega_{\rm d}}\xi \sin(\omega_{\rm d} t)\right)\hat{\sigma}_{\rm y}\right] + g(\hat{a}^{\dag}+\hat{a})\hat{\sigma}_{\rm x} + G\cos(\omega_{\rm d} t)(1-\xi)\hat{\sigma}_{\rm x}.
\end{equation}
By expanding the nested trigonometric functions with the Jacobi-Anger relations, and by neglecting the second and higher harmonics, we obtain
\begin{equation}\label{eq:S_CRHW}
\hat{H}=\omega_{\rm c} \hat{a}^{\dag}\hat{a} + \frac{\tilde{\omega}_0}{2}\hat{\sigma}_{\rm z} + \omega_0 J_1\left(\frac{2G}{\omega_{\rm d}}\xi\right)\sin(\omega_{\rm d} t)\hat{\sigma}_{\rm y} + G\cos(\omega_{\rm d} t)(1-\xi)\hat{\sigma}_{\rm x} + g(\hat{a}^{\dag}+\hat{a})\hat{\sigma}_{\rm x},
\end{equation}
where $\tilde{\omega}_0\equiv \omega_0 J_0(2G\xi/\omega_{\rm d})$ is the renormalized qubit frequency. Now, if one chooses the free parameter $\xi$ as
\begin{equation}
G(1-\xi)=\omega_0J_1\left(\frac{2G}{\omega_{\rm d}}\xi\right)\equiv \frac{\tilde{G}}{2},
\end{equation}
one is able to write the Hamiltonian into the RWA form
\begin{equation}
\hat{H}=\omega_{\rm c} \hat{a}^{\dag}\hat{a} + \frac{\tilde{\omega}_0}{2}\hat{\sigma}_{\rm z} + g(\hat{a}^{\dag}\hat{\sigma}_-+\hat{a}\hat{\sigma}_+) + \frac{\tilde{G}}{2}\left(e^{i\omega_{\rm d} t} \hat{\sigma}_- + e^{-i\omega_{\rm d} t}\hat{\sigma}_+\right).
\end{equation}
One can write the Hamiltonian in Eq.~(\ref{eq:S_Hameff}) into this same form by applying RWA for the coupling and drive terms. However, here the qubit frequency $\tilde{\omega}_0$ and the effective drive amplitude $\tilde{G}$ are renormalized by the drive. As a consequence, the cavity and the qubit are resonant for intermediate driving amplitudes and, thus, the second order Schrieffer-Wolff diagonalization becomes insufficient. However, for large drive amplitudes the qubit-cavity coupling becomes again dispersive. As shown in Fig.~2(a) of the main text, there exists a mismatch with the numerical results at high drive powers indicating the insufficiency of the RWA made by neglecting the second and higher order harmonics in Eq.~(\ref{eq:S_CRHW}).

\section{Classical model}

In this section we briefly present the classical analysis of driven qubit-resonator system.
We treat the qubit as a classical Josephson junction, which allows us to approximate it as a nonlinear inductance
\begin{equation}\label{S_RSJ}
L_{\rm J}^{-1}(\varphi_{\rm J})=\left(\frac{2e}{\hbar}\right)^2 E_{\rm J} \cos(\varphi_{\rm J}),
\end{equation}
where $\varphi_{\rm J}=(2\pi /\Phi_{0})\Phi_{\rm J}$ is the superconducting phase difference across the junction. If the driving frequency is detuned from the
resonance sufficiently far, one can split the Josephson phase into a sum consisting of a dominating contribution $\varphi_{\rm d}$ determined by the driving, and a small correction $\varphi_{\rm p}$ due to the probe tone:
\begin{eqnarray}
\varphi_{\rm J}(t) = \varphi_{\rm d}(t) + \varphi_{\rm p}(t),\;\;
\varphi_{\rm d}(t) = \varphi_{\rm d}\cos[\omega_{\rm d} t+\alpha],
\end{eqnarray}
where $\alpha$ is a phase shift, and the amplitude of phase oscillations is defined as
\begin{eqnarray}
\varphi_{\rm d}
= \sqrt{\frac{P_{\rm in}}{P_0}}.
\end{eqnarray}
In this equation $P_{\rm in}$ is the incoming microwave power and $P_0$ is the value of the driving power at which
the amplitude of phase oscillations equals to 1.

Performing the averaging of Eq. (\ref{S_RSJ}) over one period of the driving signal, and taking
the linear approximation in the small phase $\varphi_{\rm p}$, we can write the Josephson inductance as
\begin{equation}\label{S_RSJ1}
L_{\rm J}^{-1}(\varphi_{\rm d})= \left(\frac{2e}{\hbar}\right)^2 E_{\rm J} J_0(\varphi_{\rm d}),
\end{equation}
where $J_0$ is the Bessel function. Thus, the effective Josephson inductance
oscillates with the applied power.

The impedance of the circuit can be written into the form
\begin{equation}\label{eq:S_imp}
\frac{1}{Z(\omega)}=i\omega C_{\rm r}+\frac{1}{i\omega L_{\rm r}} + \frac{1}{\frac{1}{i\omega C_{\rm g}} + \frac{1}{i\omega C_{\rm J}+ \frac{1}{i\omega L_{\rm J}}}},
\end{equation}
where $C_{\rm J}$ is the capacitance of the transmon qubit.
If $L_{\rm J}\rightarrow \infty$, we have a bare resonator cavity with the eigenfrequency $\omega_{\rm c}=1/\sqrt{L_{\rm r}C_{\rm r}^{\rm eff}}$, where
\begin{equation}
C_{\rm r}^{\rm eff} = C_{\rm r}+\frac{C_{\rm g}C_{\rm J}}{C_{\rm g}+C_{\rm J}}.
\end{equation}
Similarly, if $L_{\rm r}\rightarrow \infty$, we have bare Josephson resonator with the plasma frequency $\Omega(\varphi_{\rm d})=1/\sqrt{L_{\rm J}(\varphi_{\rm d})C_{\rm J}^{\rm eff}}$, where
\begin{equation}
C_{\rm J}^{\rm eff} = C_{\rm J}+\frac{C_{\rm g}C_{\rm r}}{C_{\rm g}+C_{\rm r}}.
\end{equation}
Thus, the impedance in Eq.~(\ref{eq:S_imp}) describes two capacitively coupled harmonic $LC$-resonators.

From $1/Z(\omega) = 0$ we obtain the classical normal modes
\begin{eqnarray}
\omega_{\pm} = \sqrt{\frac{\Omega^2(\varphi_{\rm d})+\omega_{\rm c}^2}{2}
\pm\frac{1}{2}\sqrt{\left[\Omega^2(\varphi_{\rm d})-\omega_{\rm c}^2\right]^2+4B^2\Omega^2(\varphi_{\rm d})\omega_{\rm c}^2} },
\label{S_omega121}
\end{eqnarray}
where
\begin{equation}
B=\frac{C_{\rm g}}{\sqrt{(C_{\rm g}+C_{\rm J})(C_{\rm g}+C_{\rm r})}}.
\end{equation}

Then, we assume that $\Omega(0)\gg \omega_{\rm c}$ and write the lower normal mode in the low power limit ($\varphi_{\rm d}\approx 0$) as
\begin{equation}
\omega_-\approx \omega_{\rm c} - \frac{g^2}{\omega_0-\omega_{\rm c}}- \frac{g^2}{\omega_0+\omega_{\rm c}},
\end{equation}
where we have identified $\Omega^2(\varphi_{\rm d}) = \omega_0^2 J_0(\varphi_{\rm d})$ and $g=B \sqrt{\omega_{\rm c}\Omega(0)}/2$ so that at the low power limit the classical calculation matches with the quantum result obtained with the Rabi Hamiltonian. The first correction to the cavity frequency $\omega_{\rm c}$ is the dispersive vacuum Stark shift and the second term is the vacuum Bloch--Siegert correction. In terms of the new variables, the normal modes can be written as
\begin{equation}
\omega_{\pm} = \sqrt{\frac{\Omega^2(\varphi_{\rm d})+\omega_{\rm c}^2}{2}
\pm\frac{1}{2}\sqrt{\left[\Omega^2(\varphi_{\rm d})-\omega_{\rm c}^2\right]^2+16g^2\Omega^2(\varphi_{\rm d})\omega_{\rm c}/\omega_0} },\tag{3}
\end{equation}

\begin{figure}[!ht]
\includegraphics[width=0.8\linewidth]{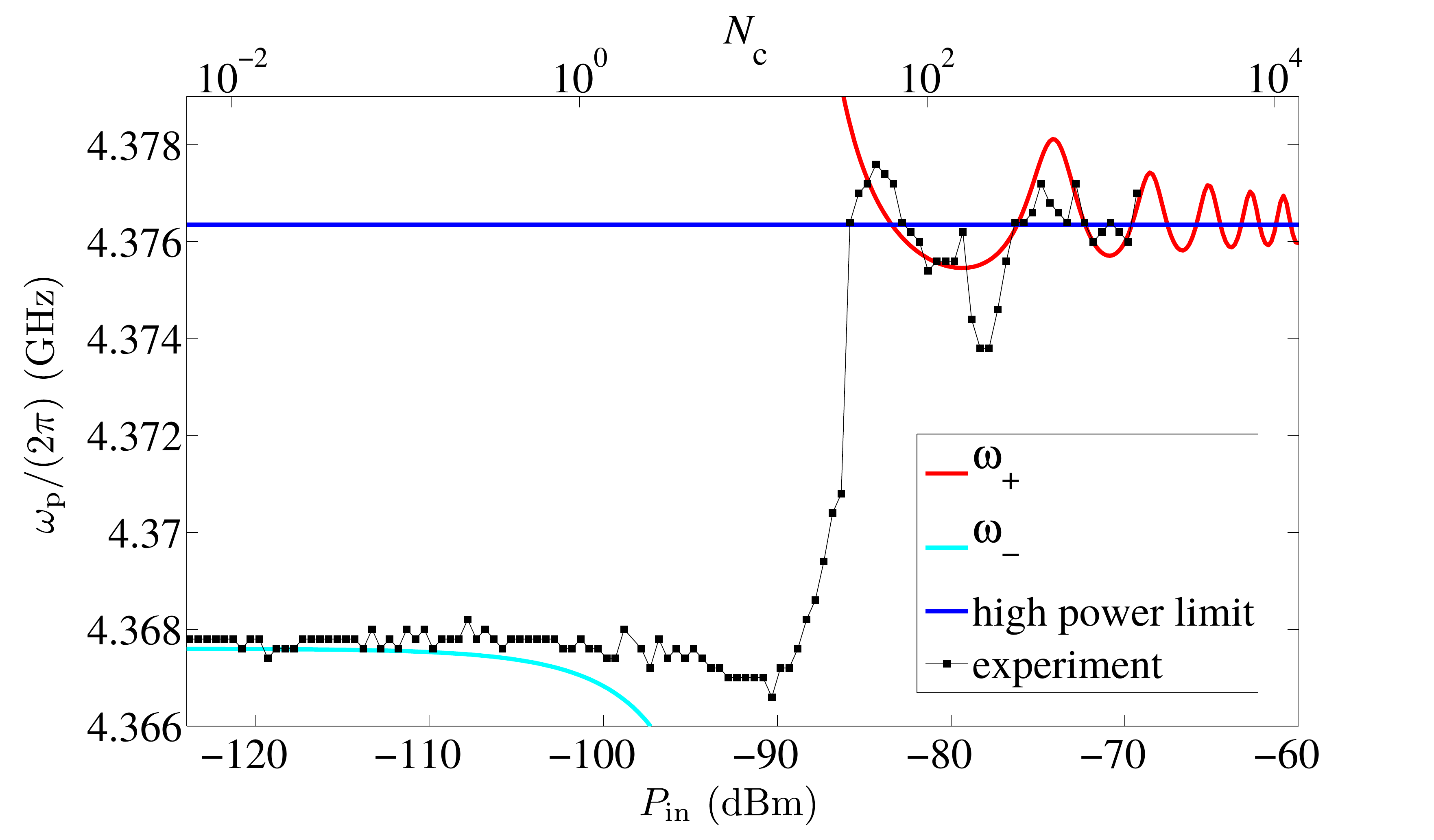}
\caption{Frequencies $\omega_\pm$, given by the Eq.~(\ref{S_omega121}), versus the applied power.
The parameters were chosen as follows: resonator frequency $\omega_{\rm c}/(2\pi) = 4.376$ GHz,
Josephson plasma frequency $\omega_0/(2\pi) = 5.160$ GHz, and the coupling frequency $g/(2\pi)=80$ MHz.}
\label{Fig:omega12}
\end{figure}

The classical normal modes $\omega_-$ and $\omega_+$ are plotted in Fig.~\ref{Fig:omega12} as a function of the applied microwave power. We have used the same values as in the simulations for the quantum regime: $\omega_{\rm c}/(2\pi) = 4.376$ GHz, $\omega_0/(2\pi)= 5.16$ GHz, and $g/(2\pi) = 80$ MHz.

\section{Power dependence of the relaxation time of the qubit}

In this section we propose a simple classical model, which explains the power dependence of the
qubit relaxation time seen in Fig.~3(b) in the main text. We treat the transmon as a classical Josephson junction with a sinusoidal
current-phase relation. Its frequency depends on the energy as follows
\begin{eqnarray}
\omega_{\rm q}(E_{\rm q}) = \omega_0 - \frac{E_{\rm C}E_{\rm q}}{\hbar^2\omega_0}, 
\label{S_omega_q_sup}
\end{eqnarray}
where $\hbar\omega_0=\sqrt{8E_{\rm J}E_{\rm C}}$ is the qubit frequency in the limit of zero energy, $E_{\rm C}=e^2/2C_\Sigma$ is the charging energy, and
\begin{eqnarray}
E_{\rm q} = \frac{\hbar^2C_\Sigma}{2}\frac{\dot\varphi_{\rm J}^2}{4e^2} + \frac{\hbar I_{\rm C}}{2e}(1-\cos\varphi_{\rm J})
\label{S_Eq_sup}
\end{eqnarray} 
is the total classical energy of the Josephson junction. For clarity, we remind the reader that
the energy dependence of the frequency arises form the anharmonicity of the qubit.
Indeed,  the dynamics of the Josephson junction is described by the equation of pendulum,
\begin{eqnarray}
\hbar C_\Sigma\frac{\ddot\varphi_{\rm J}}{2e} + I_{\rm C}\sin\varphi_{\rm J} = 0.
\end{eqnarray}
From this equation one can derive the exact formula for the frequency
\begin{eqnarray}
\omega_{\rm q}(E_{\rm q}) = \frac{\pi\omega_0}{\sqrt{2}} 
\left(\int_{0}^{\arccos(1- 8E_{\rm C}E_{\rm q}/(\hbar\omega_0)^2)}
\frac{d\varphi_{\rm J}}{\sqrt{\cos\varphi_{\rm J} - 1 + 8E_{\rm C}E_{\rm q}/(\hbar\omega_0)^2}} \right)^{-1},
\end{eqnarray}
which in the limit $E_{\rm q} \lesssim (\hbar\omega_0)^2/8E_{\rm C}$ reduces to Eq.~(\ref{S_omega_q_sup}).

If the qubit is fully isolated from the environment, its energy in Eq.~(\ref{S_Eq_sup}) does not depend on time.
However, in the experiment the qubit is weakly coupled to the resonator and to other environmental degrees of freedom.
Therefore the energy $E_{\rm q}$ may slowly change in time and its evolution is governed by the equation
\begin{eqnarray} 
\frac{dE_{\rm q}}{dt}= -\frac{E_{\rm q}-E_{\rm q}^{\rm eq}}{T_1(0)} + P(E_{\rm q}).
\label{S_Eq_Eq}
\end{eqnarray}
Here $T_1(0)$ and $E_{\rm q}^{\rm eq}$ are, respectively, the relaxation time and the equilibrium energy of the qubit in the absence of the microwave driving, and
\begin{eqnarray}
P(E_{\rm q}) \approx A\frac{ g^2\omega_{\rm c}\omega_0 }{T_1(0)\left[\left(\omega_{\rm q}^2(E_{\rm q})-\omega_{\rm c}^2\right)^2 + \omega_{\rm c}^2/T_{1}^{2}(0) \right]}
\frac{\omega_{\rm d}\omega_{\rm c}\kappa}{\left(\omega_{\rm d}^2-\omega_{\rm c}^2\right)^2 + \kappa^2\omega_{\rm d}^2} P_{\rm in}
\label{S_PEq}
\end{eqnarray}
is the fraction of the driving microwave power which is absorbed in the qubit. The precise value of the constant pre-factor $A$ will not be important here.
The expression~(\ref{S_PEq}) has a simple physical meaning: in order to be absorbed,  
the incoming microwave power
has to pass through the resonator and then to excite the qubit; at both stages it is multiplied
by a frequency dependent filtering factor corresponding to the response function of a damped harmonic oscillator and having a Lorentzian shape.

In the stationary case one should put $dE_{\rm q}/dt=0$ in Eq.~(\ref{S_Eq_Eq}). Afterwards it is reduced
to the equation for the stationary value of the qubit energy, $E_{\rm q}^{\rm st}$, in the presence of pumping,
\begin{eqnarray}
E_{\rm q}^{\rm st} = E_{\rm q}^{\rm eq} + T_1(0)P(E_{\rm q}^{\rm st}).
\label{S_Est}
\end{eqnarray} 
Applying a weak excitation signal directly to the qubit, we slightly move the energy away from the stationary value.
After this signal is switched off, the energy relaxes back to $E_{\rm q}^{\rm st}$ accoring to Eq.~(\ref{S_Eq_Eq}) linearized around $E_{\rm q}^{\rm st}$,
\begin{eqnarray}
\frac{d}{dt}\delta E_{\rm q} = - \frac{\delta E_{\rm q}}{T_1(0)} + \frac{dP(E_{\rm q}^{\rm st})}{d E_{\rm q}}\delta E_{\rm q}.
\end{eqnarray}
Here the energy deviation  $\delta E_{\rm q} = E_{\rm q}-E_{\rm q}^{\rm st}$ is supposed to be small, $|\delta E_{\rm q}|\ll |\hbar\omega_0-\hbar\omega_{\rm c}|\hbar\omega_0/E_{\rm C} $. Hence the power dependent  relaxation time, $T_1(P_{\rm in})$, is given by the expression
\begin{eqnarray}
\frac{1}{T_1(P_{\rm in})} = \frac{1}{T_1(0)} - \frac{dP(E_{\rm q}^{\rm st})}{d E_{\rm q}}.
\end{eqnarray}
This simple mechanism of change of relaxation time in a pumped device has also been studied in the theory of superconducting transition-edge radiation sensors. 
In that case negative electrothermal feedback makes the response time of the device significantly shorter~\cite{S_Irwin,S_Irwin2}.

Evaluating the derivative $dP/dE_{\rm q}$ from Eq.~(\ref{S_PEq}) and assuming that $\omega_{\rm c}/T_1(0)\ll 1$, we find
\begin{eqnarray}
\frac{1}{T_1(P_{\rm in})} = \frac{1}{T_1(0)} + \frac{4\omega_{\rm q}(E_{\rm q}^{\rm st})P(E_{\rm q}^{\rm st})}{\omega_{\rm q}^2(E_{\rm q}^{\rm st})-\omega_{\rm c}^2} \frac{d\omega_{\rm q}}{dE_{\rm q}}
= \frac{1}{T_1(0)} - \frac{4\omega_{\rm q}(E_{\rm q}^{\rm st})P(E_{\rm q}^{\rm st})}{\omega_{\rm q}^2(E_{\rm q}^{\rm st})-\omega_{\rm c}^2} \frac{E_{\rm C}}{\hbar^2\omega_0},
\label{S_T1P}
\end{eqnarray}
where we have used  Eq.~(\ref{S_omega_q_sup}) in order to find the derivative $d\omega_{\rm q}/dE_{\rm q}$.
Next, we exclude the power $P(E_{\rm q}^{\rm st})$, which is not directly measurable, from this expression. From Eq.~(\ref{S_Est})
we find $P(E_{\rm q}^{\rm st}) = (E_{\rm q}^{\rm st}-E_{\rm q}^{\rm eq})/T_1(0)$, 
while from Eq.~(\ref{S_omega_q_sup}) we get $E_{\rm q}^{\rm st}-E_{\rm q}^{\rm eq}=(\hbar\omega_{\rm q}(E_{\rm q}^{\rm eq})-\hbar\omega_{\rm q}(E_{\rm q}^{\rm st}))\hbar\omega_0/E_{\rm C}$.
Combining the two expressions we write the power in terms of the measurable power dependent qubit frequency $\omega_{\rm q}(P_{\rm in})\equiv \omega_{\rm q}(E_{\rm q}^{\rm st})$, and find
$P(E_{\rm q}^{\rm st}) = (\hbar\omega_{\rm q}(0)-\hbar\omega_{\rm q}(P_{\rm in}))\hbar\omega_0/E_{\rm C}T_1(0)$. Substituting this expression in the Eq.~(\ref{S_T1P}), we arrive at the final expression for the relaxation time of the qubit expressed in terms of measurable parameters
\begin{eqnarray}
\frac{1}{T_1(P_{\rm in})} 
= \frac{1}{T_1(0)}\left[1 - \frac{4\omega_{\rm q}(0)\big[\omega_{\rm q}(0)-\omega_{\rm q}(P_{\rm in})\big]}{\omega_{\rm q}^2(P_{\rm in})-\omega_{\rm c}^2} \right].
\label{S_T1Pfinal}
\end{eqnarray}
Eq.~(\ref{S_T1Pfinal}) is valid provided the frequency of the resonator does not depend on the driving power, which is the case in our experiment only in a certain range of powers, corresponding to $N_{c}\approx 2$. 

\section{Simulation for a multi-state transmon}

\subsection{Floquet theorem}
The Hamiltonian in Eq.~(\ref{eq:S_Hameffmulti}) is $\tau\equiv 2\pi/\omega_{\rm d}$-periodic which allows the use of the Floquet theorem~\cite{S_Shirley65}. The theorem states that the solutions to the time-dependent Schr\"odinger equation are of the form $\vert\Psi(t)\rangle = e^{-i\varepsilon t}\vert\Phi(t)\rangle$,
where $\varepsilon$ are the quasienergies and $\vert\Phi(t)\rangle$ are the corresponding $\tau$-periodic quasienergy states.

Due to the $\tau$-periodicity of $\vert\Phi(t)\rangle$ and $\hat{H}(t)$ we can write them in terms of Fourier series
\begin{eqnarray}
\vert\Phi(t)\rangle &=& \sum_{n=-\infty}^\infty e^{in\omega_{\rm d} t}\vert\Phi^{(n)}\rangle ,\\
\hat{H}(t) &=& \sum_{n=-\infty}^\infty e^{in\omega_{\rm d} t} \hat{H}^{(n)}\,.
\end{eqnarray}
Plugging these into the Schr\"odinger equation we get the $m^{\text{th}}$ Fourier component as
\begin{equation}
\sum_n [n\omega_{\rm d}\delta_{nm} +\hat{H}^{(m-n)}]\vert\Phi^{(n)}\rangle = \varepsilon\vert\Phi^{(m)}\rangle\,.
\end{equation}
We can write this as a time-independent eigenvalue problem
\begin{equation}
\hat{H}_{\rm F} \vert\Phi\rangle = \varepsilon\vert\Phi\rangle\,,
\label{eq:S_Floquet}
\end{equation}
where
\begin{equation}
\langle \sigma,N,n\vert \hat{H}_{\rm F}\vert\sigma',M,m\rangle = n\omega_{\rm d}\delta_{\sigma,\sigma'} \delta_{NM} \delta_{nm} +\frac{1}{\tau}\int_0^\tau \text{d}t\, e^{-i(n-m)\omega_{\rm d} t} \langle\sigma, N\vert \hat{H}(t)\vert \sigma',M\rangle
\end{equation}
and
\begin{equation}
\langle\sigma,N,n\vert\Phi\rangle = \frac{1}{\tau}\int_0^\tau \text{d}t\, e^{-in\omega_{\rm d} t} \langle\sigma,N\vert \Phi(t)\rangle\,.
\end{equation}
As a solution to the Floquet eigenvalue equation~(\ref{eq:S_Floquet}), we obtain the quasienergies $\varepsilon_{\alpha, \ell}= \varepsilon_{\alpha} + \ell \omega_{\rm d}$ and the corresponding eigenstates $|\alpha,\ell\rangle \equiv |\Phi_{\alpha,\ell}(t)\rangle = e^{i\ell\omega_{\rm d} t} |\Phi_{\alpha,0}(t)\rangle$.

\subsection{Absorption rate and connection to reflection measurement}

In the reflection measurement, one measures the reflection coefficient $\Gamma$ of the weak probe. When the response is linear, the coefficient can be written into the form
\begin{equation}
\Gamma(\omega_{\rm p})=\frac{Z(\omega_{\rm p})-Z_0}{Z(\omega_{\rm p})+Z_0},
\end{equation}
where $Z_0\approx 50 \ \Omega$ is the impedance of the transmission line, and
\begin{equation}
Z(\omega) = \frac{1}{i\omega C_c}+ Z_{\rm S}(\omega),
\end{equation}
with $Z_{\rm S}(\omega)=Z'(\omega)+iZ''(\omega)$ as the impedance of the strongly driven transmon-cavity system. In our setup, the weak probe is of the form
\begin{equation}
\hat{H}_{\rm P}(t) = A_{\rm p} \cos(\omega_{\rm p} t) (\hat{a}^{\dag}+\hat{a}).
\end{equation}
We calculate the impedance by using Fermi's golden rule for probe induced transition between the quasienergy states. We obtain
\begin{equation}
\begin{split}
Z'(\omega_{\rm p}) &=\frac{Z_{\rm c}}{4}\sum_{i,f,\ell} (\omega_{fi}+\ell\omega_{\rm d}) p_i \gamma_{fi}|\langle f,\ell|(\hat{a}^{\dag}+\hat{a})|i,0\rangle|^2\left[\frac{1}{(\omega_{fi} + \ell \omega_{\rm d}-\omega_{\rm p})^2+\frac14\gamma_{fi}^2}+\frac{1}{(\omega_{fi} + \ell \omega_{\rm d}+\omega_{\rm p})^2+\frac14\gamma_{fi}^2}\right],
\end{split}
\end{equation}
where $Z_{\rm c}=\sqrt{L_{\rm c}/C_{\rm c}}$, $\omega_{fi} = \varepsilon_i -\varepsilon_f$, $p_i$ is the initial occupation probability of state $|i,0\rangle$ and $\gamma_{fi}$ is the relaxation rate from state $|f,0\rangle$ to $|i,0\rangle$. The first term in the second line gives the emission to and the latter absorption from the probe. The imaginary part $Z''(\omega)$ of the impedance is given by the real part in terms of the Kramers-Kronig relation~\cite{S_LL}:
\begin{equation}
Z''(\omega_{\rm p})= \frac{1}{\pi}\mathrm{P} \int_{-\infty}^{\infty} \frac{Z'(\xi)}{\xi-\omega_{\rm p}}d\xi = \frac{Z_{\rm c}}{4}\sum_{i,f,\ell}p_i |\langle f,\ell|(\hat{a}^{\dag}+\hat{a})|i,0\rangle|^2\frac{4\omega_{\rm p}(\omega_{fi}+\ell\omega_{\rm d})\left[(\omega_{fi}+\ell\omega_{\rm d}-\omega_{\rm p})(\omega_{fi}+\ell\omega_{\rm d}+\omega_{\rm p}) - \frac14\gamma_{fi}^2)\right]}{[(\omega_{fi}+\ell\omega_{\rm d}-\omega_{\rm p})^2+\frac14\gamma_{fi}^2][(\omega_{fi}+\ell\omega_{\rm d}+\omega_{\rm p})^2+\frac14\gamma_{fi}^2]},
\end{equation}
where P stands for principal value. Thus, the reflection coefficient in the linear response can be readily obtained once the equilibrium occupations $p_i$ and the transition rates $\gamma_{fi}$ between the quasienergy states have been solved.

\subsection{Dissipation in a strongly driven system}

In order to calculate $Z_{\rm S}(\omega)$, we need to find out $p_i$ and $\gamma_{fi}$. Let us consider a strongly driven system coupled to a bosonic bath formed by a set of noninteracting harmonic oscillators:
\begin{equation}
\hat{H}(t)= \hat{H}_{\rm S}(t) + \hat{H}_{\rm B} + \hat{H}_{\rm SB},
\end{equation}
where $\hat{H}_{\rm S}(t)$,
\begin{equation}
\hat{H}_{\rm B}=\sum_i \omega_i \hat{a}^{\dag}_i\hat{a}_i,
\end{equation}
and
\begin{equation}
\hat{H}_{\rm SB}= -\hat{X}\sum_i \frac{g_i}{\sqrt{2C_i\omega_i}} (\hat{a}_i +\hat{a}_i^\dagger) + \hat{X}^2 \sum_i \frac{g_i^2}{2C_i\omega_i^2}
\end{equation}
are the Hamiltonians of the driven system, the bath and the coupling, respectively. Here the bath is described by a set of harmonic oscillators, and $\hat{a}_i$, $C_i$ and $\omega_i$ are the annihilation operator, capacitance and angular frequency of the $i$th bath oscillator, respectively.

The bath is coupled to the system via the operator $\hat{X}$. In our case, $\hat{X}=\frac{1}{\sqrt{2C_{\rm r}\omega_{\rm c}}}(\hat{a}^{\dag}+\hat{a})$, where $C_{\rm r}$ is the capacitance of the cavity. Proceeding similarly as in Refs.~\cite{S_Blumel91} and~\cite{S_Grifoni98} we obtain

\begin{equation}
\begin{split}
p_{\alpha} &= \frac{\sum_{\nu\neq \alpha} \Gamma_{\nu\alpha}p_{\nu}}{\sum_{\nu\neq \alpha}\Gamma_{\alpha\nu}}\\
\gamma_{\alpha\beta} &\equiv \frac12 \sum_{\nu} \left[\Gamma_{\alpha\nu}+\Gamma_{\beta\nu}\right]\\
\Gamma_{\alpha\beta}&=\sum_{\ell=-\infty}^{\infty} \left[\gamma_{\alpha\beta \ell}+n_{\rm th}(|\Delta_{\alpha\beta \ell}|)\left(\gamma_{\alpha\beta \ell}+\gamma_{\beta \alpha -\ell}\right)\right]\\
\gamma_{\alpha\beta\ell} &= \frac{\pi}{2} \kappa \theta(\Delta_{\alpha\beta\ell})\frac{\Delta_{\alpha\beta\ell}}{\omega_{\rm c}}|\langle \alpha, \ell|(\hat{a}^{\dag}+\hat{a})|\beta,0\rangle|^2,
\end{split}
\end{equation}
where $\theta(\omega)$ is the Heaviside step-function, $n_{\rm th}(\omega)=1/(e^{\hbar\omega/(k_{\rm B} T)}-1)$ is the Bose-Einstein occupation in the environment at temperature $T$, $\Delta_{\alpha \beta \ell} = \varepsilon_{\alpha} - \varepsilon_{\beta} + \ell\omega_{\rm d}$ and $\ket{\alpha,\ell}$ are the Floquet states corresponding to energy $\varepsilon_\alpha +\ell\omega_{\rm d}$.

For a fixed $\ell$, we see that the transition rates obey detailed balance. However, the resulting transition rates $\Gamma_{\alpha\beta}$ do not obey the detailed balance (see also Ref.~\cite{S_Gasparinetti13}).

\section{Additional experimental data}
\begin{figure}[h]
\centering
\includegraphics[width=1.0\linewidth]{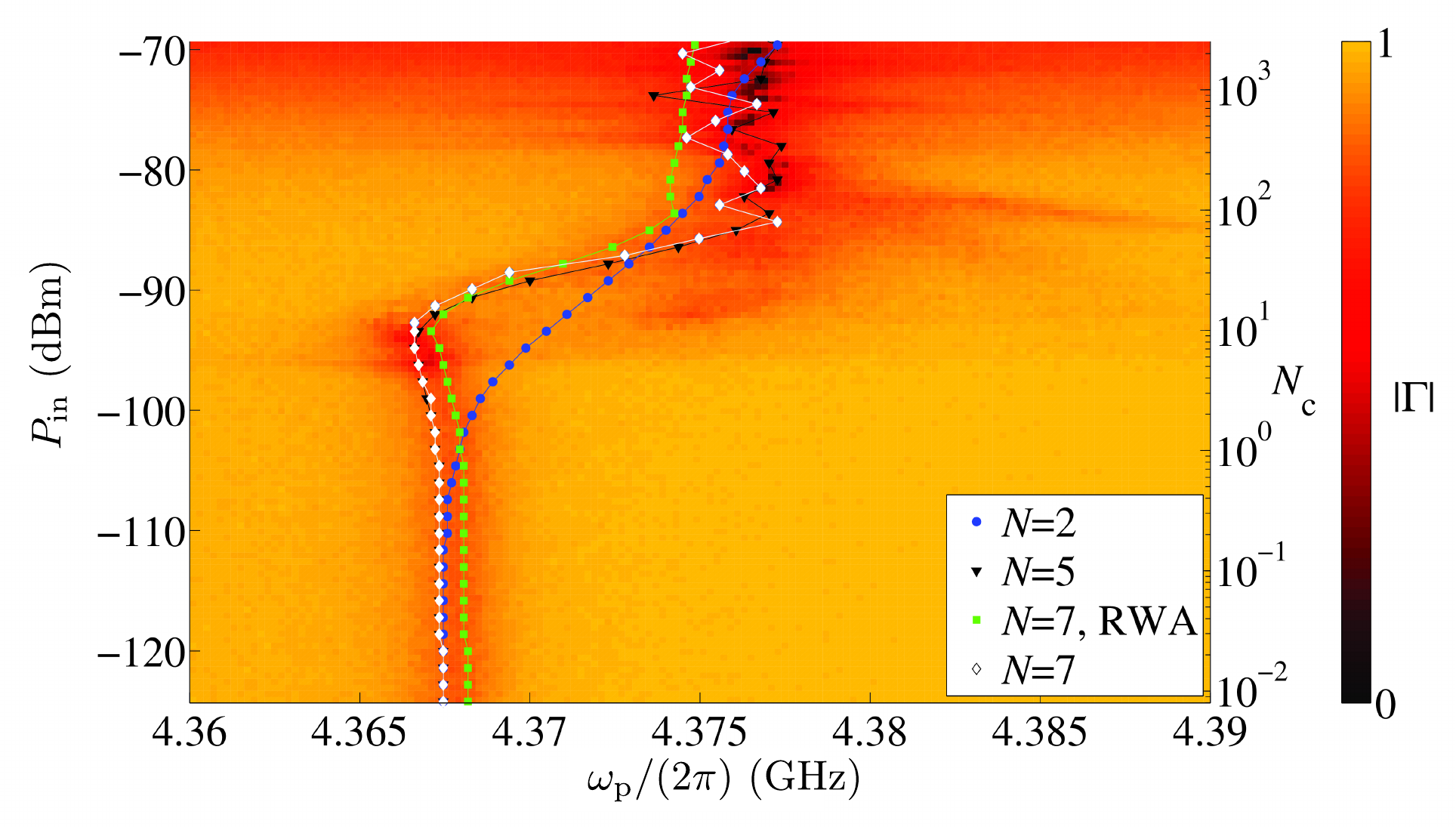}
\caption{Measured reflection coefficient $|\Gamma|$ shown as a function of the probe frequency $\omega_{\rm p}$ and the input power $P_{\rm in}$ (average number of cavity quanta $N_{\rm c}$). The numerical data are shown with solid line contours. The drive frequency $\omega_{\rm d}/(2\pi) = 4.355$ GHz and other parameters are the same as in the main text.}\label{fig:expBSshift2}
\end{figure}

We have also measured the reflection coefficient with different drive frequencies. In Fig.~\ref{fig:expBSshift2} we have the comparison between the measured reflection coefficient and the numerical simulations with the drive frequency $\omega_{\rm d}/(2\pi)=4.355$ GHz. Similarly as in the Fig.~3 of the main paper, we can see how the simulations converge towards the experimental result as more transmon states are included.

\begin{figure}
  \centering
    \begin{subfigure}[]{0.3\textwidth}
      \centering

\includegraphics[width=1\linewidth]{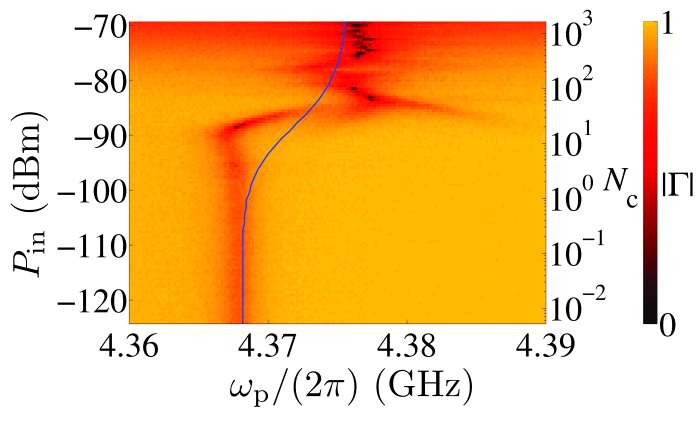}
      \captionof{figure}{$\omega_{\rm d}/(2\pi)=4.350$ GHz}
      \label{fig:s1wdSweep1}
    \end{subfigure}%
    \begin{subfigure}[]{0.3\textwidth}
      \centering

\includegraphics[width=1\linewidth]{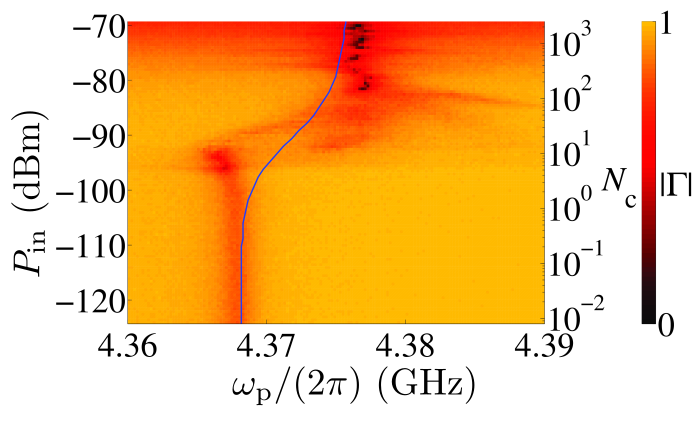}
      \captionof{figure}{$\omega_{\rm d}/(2\pi)=4.355$ GHz}
      \label{fig:s1wdSweep2}
    \end{subfigure}
        \begin{subfigure}[]{0.3\textwidth}
      \centering

\includegraphics[width=1\linewidth]{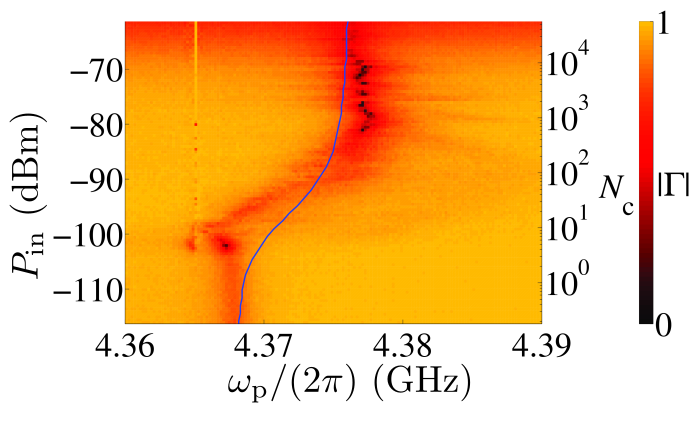}
      \captionof{figure}{$\omega_{\rm d}/(2\pi)=4.365$ GHz}
      \label{fig:s1wdSweep3}
    \end{subfigure}

    \begin{subfigure}[]{0.3\textwidth}
      \centering

\includegraphics[width=1\linewidth]{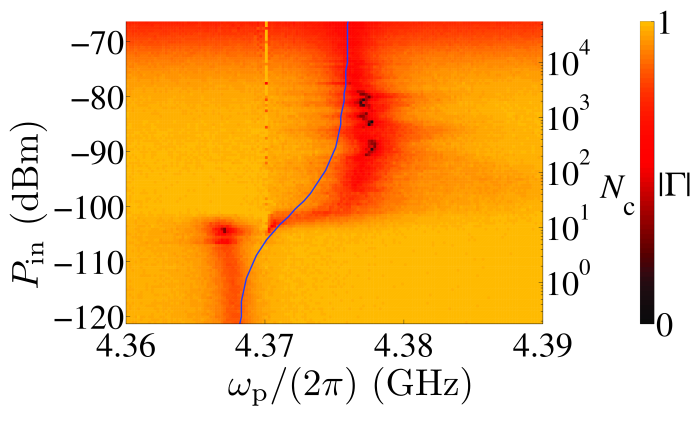}
      \captionof{figure}{$\omega_{\rm d}/(2\pi)=4.370$ GHz}
      \label{fig:s1wdSweep4}
    \end{subfigure}
    \begin{subfigure}[]{0.3\textwidth}
      \centering

\includegraphics[width=1\linewidth]{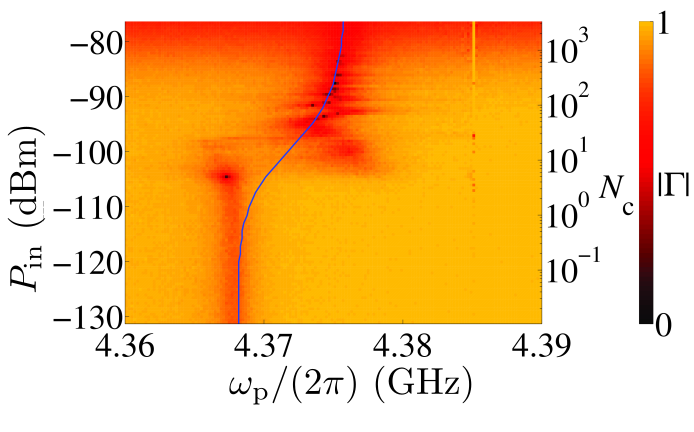}
      \captionof{figure}{$\omega_{\rm d}/(2\pi)=4.385$ GHz}
      \label{fig:s1wdSweep5}
    \end{subfigure}
    \begin{subfigure}[]{0.3\textwidth}
      \centering

\includegraphics[width=1\linewidth]{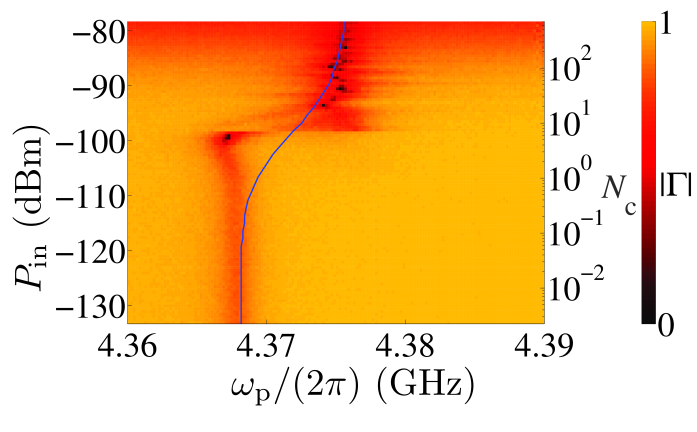}
      \captionof{figure}{$\omega_{\rm d}/(2\pi)=4.390$ GHz}
      \label{fig:s1wdSweep6}
    \end{subfigure}
  \caption{Measured reflection coefficient $|\Gamma|$ as a function of the probe frequency $\omega_{\rm p}$ and the power $P_\text{in}$ (average number of cavity quanta $N_{\rm c}$). The driving tone is also visible in (c) (d) and (e) as a narrow vertical line. The blue lines represent the simulated two-level RWA resonance.}
  \label{fig:s1wdSweep}
\end{figure}

\begin{figure}
  \centering
    \begin{subfigure}[]{0.3\textwidth}
      \centering

\includegraphics[width=1\linewidth]{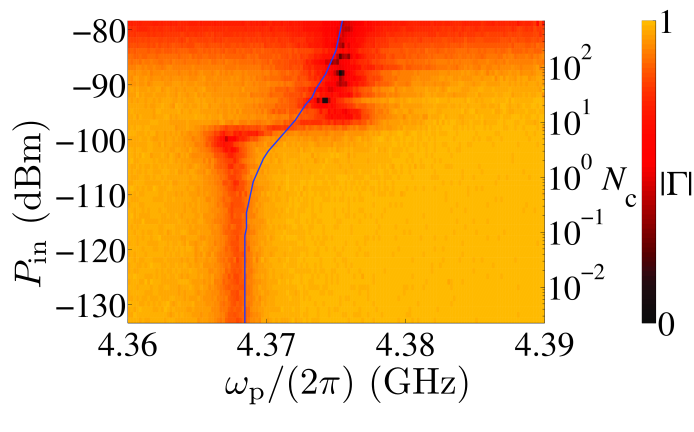}
      \captionof{figure}{$\omega_0/(2\pi)=5.188$ GHz}
      \label{fig:s1wqSweep1}
    \end{subfigure}%
    \begin{subfigure}[]{0.3\textwidth}
      \centering

\includegraphics[width=1\linewidth]{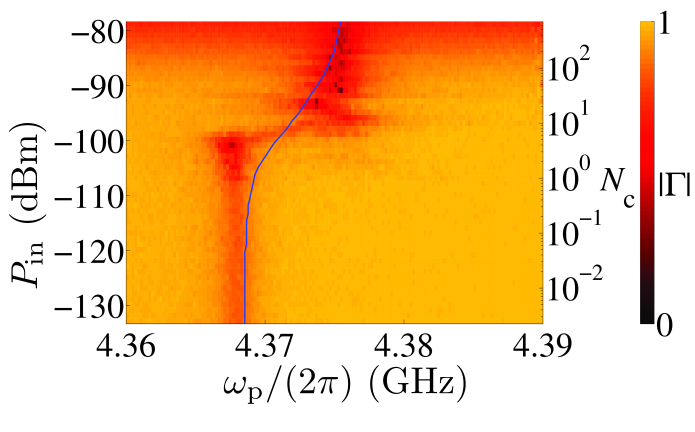}
      \captionof{figure}{$\omega_0/(2\pi)=5.208$ GHz}
      \label{fig:s1wqSweep2}
    \end{subfigure}
        \begin{subfigure}[]{0.3\textwidth}
      \centering

\includegraphics[width=1\linewidth]{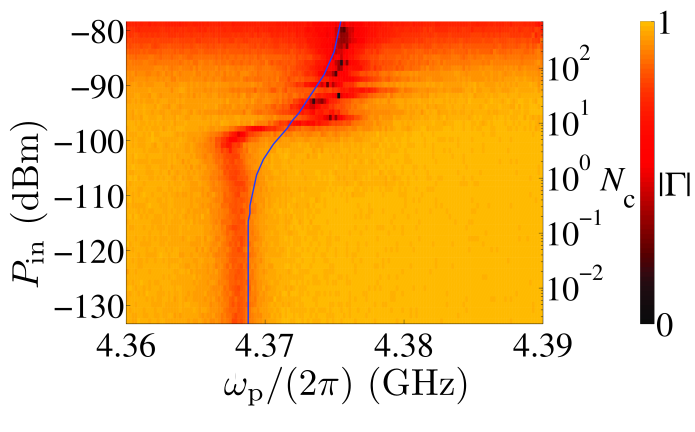}
      \captionof{figure}{$\omega_0/(2\pi)=5.226$ GHz}
      \label{fig:s1wqSweep3}
    \end{subfigure}

    \begin{subfigure}[]{0.3\textwidth}
      \centering

\includegraphics[width=1\linewidth]{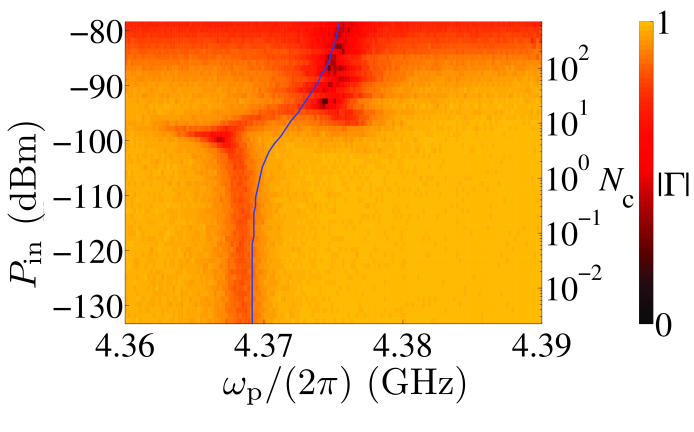}
      \captionof{figure}{$\omega_0/(2\pi)=5.282$ GHz}
      \label{fig:s1wqSweep4}
    \end{subfigure}
    \begin{subfigure}[]{0.3\textwidth}
      \centering

\includegraphics[width=1\linewidth]{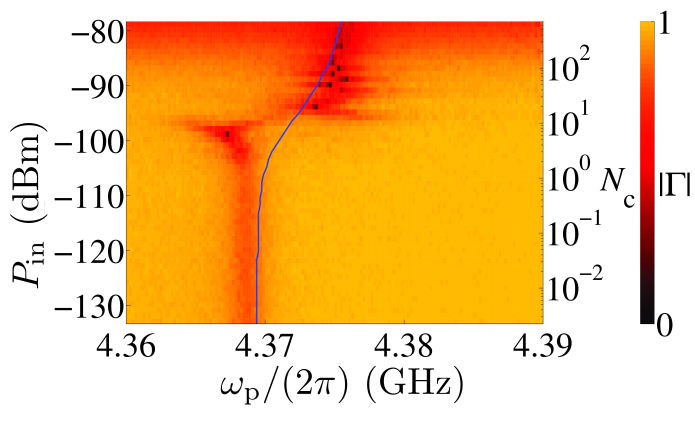}
      \captionof{figure}{$\omega_0/(2\pi)=5.318$ GHz}
      \label{fig:s1wqSweep5}
    \end{subfigure}
    \begin{subfigure}[]{0.3\textwidth}
      \centering

\includegraphics[width=1\linewidth]{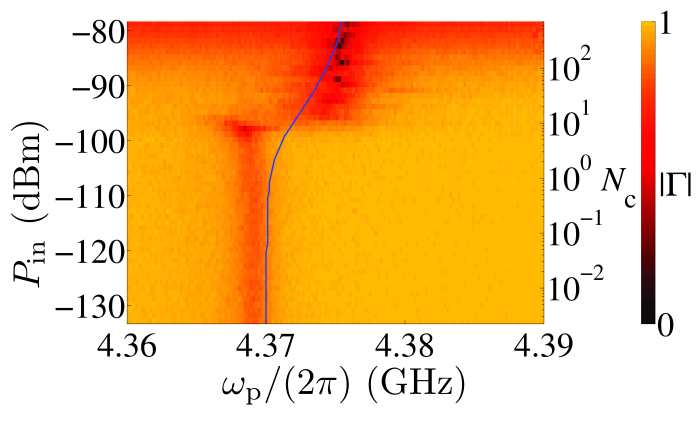}
      \captionof{figure}{$\omega_0/(2\pi)=5.414$ GHz}
      \label{fig:s1wqSweep6}
    \end{subfigure}
  \caption{Measured reflection coefficient $|\Gamma|$ as a function of the probe frequency $\omega_{\rm p}$ and the power $P_\text{in}$ (average number of cavity quanta $N_{\rm c}$). The drive frequency is $\omega_{\rm d}/(2\pi)=4.390$ GHz.  The blue lines represent the simulated two-level RWA resonance.}
  \label{fig:s1wqSweep}
\end{figure}

In Fig.~\ref{fig:s1wdSweep}, we present the measured reflection coefficient for several different drive frequencies and in Fig.~\ref{fig:s1wqSweep}, we show the data for the reflection with different qubit frequencies at fixed $\omega_{\rm d}/(2\pi)=4.390$ GHz. The numerical two-state RWA solutions are also plotted in the figures. In all figures we can see the nonmonotonic behaviour at large cavity occupations, caused by the BS effect.

\end{document}